\RequirePackage{fix-cm}
\documentclass[twocolumn]{svjour3}          
\smartqed  
\usepackage{latexsym, amssymb, amsmath,mathptmx}     
\usepackage{soul}
\usepackage{graphicx, hyperref, color}
\usepackage{esvect}
\usepackage{textcomp,marvosym}
\usepackage{diagbox}
\usepackage{xcolor, colortbl,float}
\usepackage{enumerate} 

\begin{document}
\title{Reaction time impairments in decision-making networks as a diagnostic marker for traumatic brain injuries and neurodegenerative diseases}

\titlerunning{Impaired Decision Making}        

\author{Pedro D. Maia         \and
       J. Nathan Kutz 
}


\institute{Pedro D. Maia \at
              Department of Applied Mathematics, University of Washington, \\ Seattle, WA. 98195-3925 USA. \\
              \email{pedro.doria.maia@gmail.com}           
           \and
           J. Nathan Kutz  \at
                Department of Applied Mathematics, University of Washington, \\ Seattle, WA. 98195-3925 USA.  \\
              \email{kutz@uw.edu} 
              }

\date{August 18, 2016}

\maketitle

\begin{abstract}
The presence of diffuse Focal Axonal Swellings (FAS) is a hallmark cellular feature in many neurodegenerative diseases 
and traumatic brain injury.  Among other things, the FAS have a significant impact on spike-train encodings that propagate 
through the affected neurons, leading to compromised signal processing on a neuronal network level.  This work merges, 
for the first time, three fields of study:  (i)  signal processing in excitatory-inhibitory (EI) networks of neurons via population 
codes, (ii) decision-making theory driven by the production of evidence from stimulus, and (iii) compromised spike-train 
propagation through FAS.  As such, we demonstrate a mathematical architecture capable of characterizing compromised 
decision-making driven by cellular mechanisms.  The computational model also leads to several novel predictions and 
diagnostics for understanding injury level and cognitive deficits, including a key finding that decision-making reaction times, 
rather than accuracy, are indicative of network level damage.  The results have a number of translational implications, including 
that the level of network damage can be characterized by the reaction times in simple cognitive and motor tests.
\end{abstract}


\begin{figure*}[t]
\centering{\includegraphics[width=0.8\textwidth]{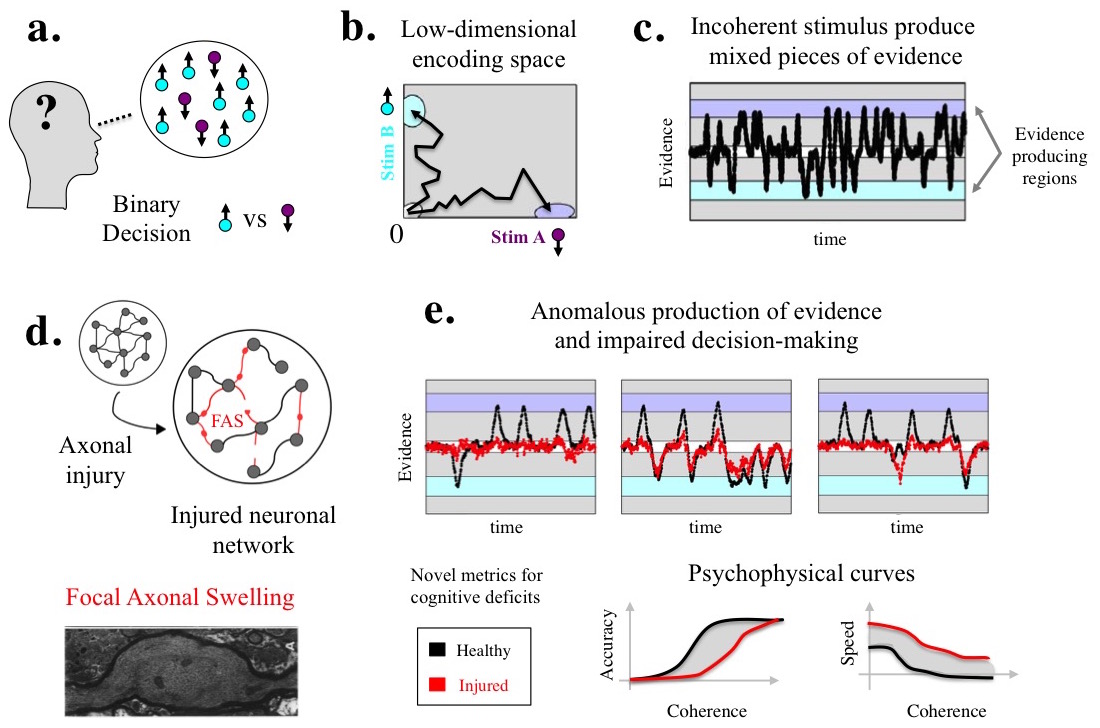}}
\caption{Schematics of the paper. 
~\textbf{a.} Initial goal is to build a system with healthy binary decision-making capabilities (A vs B) from a 
neuronal network. 
~\textbf{b.} Collective neuronal activity is mapped to a trajectory in a proper low-dimensional PCA space
(as in \cite{Laurent}). Different stimuli produce neural responses (encodings) ending at different 
fixed points (as in \cite{Shlizerman2014}). We interpret the vicinities of the fixed points as evidence-producing regions 
for the corresponding stimuli. 
~\textbf{c.} Trajectories for incoherent/mixed stimuli visit distinct evidence-producing regions multiple times. 
~\textbf{d.} We add Focal Axonal Swellings (FAS) to the network based on available experimental literature 
(see \cite{Wang2011} and Sec.~3.1), incorporating their injurious effects to spike trains 
\cite{Maia2015,Maia2014_2,Maia2014_1}. ~\textbf{e.} Virtual lesions lead to a variety of types of anomalous production 
of evidence. When integrated over time by a decision variable (as in \cite{Shadlen2013}), they culminate in interpretable 
psychophysical deficits such as accuracy and speed loss.}
\label{intro_scheme}       
\end{figure*}
%
%
%
%
%
%

\section{Introduction}
Neurodegenerative diseases and traumatic brain injuries are responsible for an overwhelming 
variety of functional deficits in animals and humans, with common developments including memory 
loss or behavioral/cognitive impairments. Many brain disorders have a complex cascade of pathological 
effects spanning multiple spatial scales: from cellular or network levels to tissues or entire brain regions. 
Our limited ability to diagnose cerebral malfunctions \textit{in vivo} cannot detect several anomalies that 
occur on smaller scales. In this sense, Focal Axonal Swellings (FAS) are ubiquitous to many incurable 
disorders such as Multiple Sclerosis, Alzheimer's and Parkinson's diseases, dramatically affecting signaling 
properties of neurons associated with cognitive tasks. Moreover, FAS are a hallmark feature of traumatic 
brain injury. To evaluate the impact of FAS in neuronal networks, we consider a computational setting where 
a healthy network produces evidence for different stimuli in the form of low-dimensional encodings (population 
codes). When presented with incoherent stimuli, the system integrates all mixed evidence, displaying stereotypical 
binary decision-making capabilities. 
The addition of axonal lesions and their impact on spike trains decrease the system's ability to transmit information 
with functional population codes. FAS also create a variety of anomalies regarding the production of evidence, which includes 
both biased and unbiased errors, confusion, and amplification of false, small signatures. 
In all cases, reaction time deficits are the most aggravating effect of FAS to decision-making performance, with reaction 
times slowing proportionally to the network's injury level. We suggest that decision-making tasks are not only a 
window to study cognition, but also a diagnostic platform for a variety of neurological pathologies and disorders.

\begin{figure*}[th]
\centering
\includegraphics[width=0.65\textwidth]{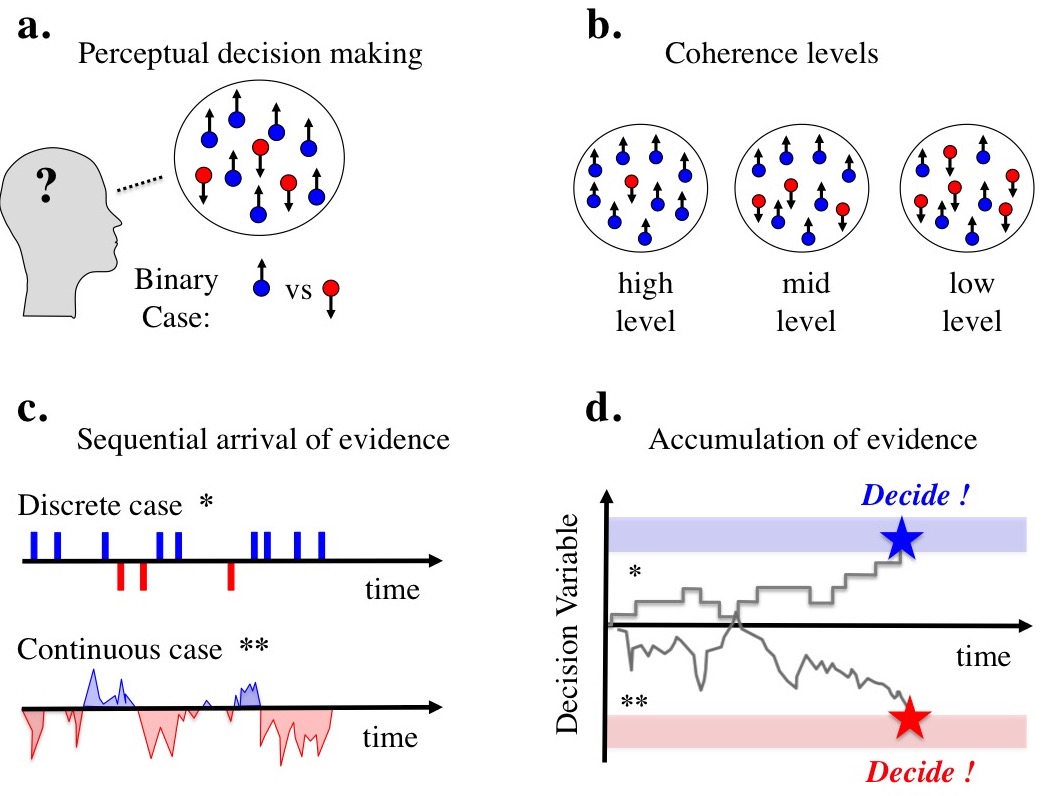}
\caption{Schematics for perceptual decision making framework. 
~\textbf{a-b.} In this work, we consider binary decisions made upon the 
sequential analysis of evidence. A common experimental setup consists 
of presenting subjects to a patch of moving dots, where some move coherently 
in a given direction (as explained in \cite{Shadlen2007}). The percentage of coherently moving 
regulates the difficulty of the task: high/low coherence levels makes the 
task easier/harder. ~\textbf{c.}  Evidence could arrive in discrete time steps or 
continuously in time. In our scheme, different signs/colors of evidence pieces are 
associated to distinct possible choices.  
~\textbf{d.}
The decision-maker continuously integrates/accumulates evidence (represented 
here by a decision variable). The process may be interrupted by committing to 
one of the alternatives, i.e., by reaching a decision-boundary associated to one 
of the alternatives. This task can be performed by both human and monkey subjects, 
allowing for both psychophysical and physiological investigation (see \cite{Shadlen2013,Shadlen2007}).
}
\label{DM_scheme}       
\end{figure*}
%
%

Figure~\ref{intro_scheme} provides a schematic overview of this 
computational study. We posit a neuronal network with biologically plausible binary Decision-Making (DM) 
capabilities (A vs B). The core of the DM system consists of an input/output network of excitatory-inhibitory (EI) neurons that 
responds to meaningful stimuli with robust, low-dimensional population codes. These codes are coupled 
to an accumulation model that integrates all produced evidence and commits to an alternative when a decision boundary 
is reached. The system's performance is described by accuracy/speed psychophysical curves as 
a function of the task's difficulty level. Throughout each step of the model, we describe a variety of  
impairments caused by the focal axonal swellings developed at a cellular level; from hindered network 
responses to the anomalous production of evidence that if accumulated over time, leads to quantifiable cognitive deficits.  

Although DM theory (see the review article~\cite{Shadlen2016} and references therein) and neural 
network computational studies (see the reviews~\cite{Network_Review,Dayan2001,Gerstner2014} 
and references therein), especially EI networks, are quite mature in the computational neuroscience community,
the spike-train propagation through damaged axons has only recently been explored~\cite{Maia2014_1,Maia2014_2,Maia2015}.
By merging these three distinct areas of study, we can provide a general theoretical architecture for
understanding how the cellular progression of neurodegenerative diseases and/or traumatic brain injury can directly lead 
to compromised DM and cognitive deficits.  Thus, by merging three key areas of study, we advance the theory linking 
cell level biophysics to behavioral decision-making capabilities.  Critical to the viability of the theory is the direct connection of 
the FAS imposed in the computational model with biophysically observed FAS statistics of swelling frequency
and size distribution.  Such statistics have been well characterized in a number of TBI studies, allowing us to calibrate
our models to state-of-the-art biophysical observations.  It is envisioned that the model can be easily integrated
with other neurodegenerative disease studies and their FAS statistics.

The paper is outlined as follows: 
In Sec.~2, we review the main ideas behind decision making (drift-diffusion) models 
and provide a detailed background on Focal Axonal Swellings (FAS). Specifically, we list several 
neurological disorders in which FAS are implicated and discuss their effects to spike train propagation. 
Section~3 describes our calibration of FAS distributions from the experimental work of Wang et al. 
\cite{Wang2011}, our network model and its population codes, along with their interpretation in a 
proper low-dimensional space. We discuss our results in Sec. ~4, which includes hindered network 
responses,  loss of transmitted information, anomalous production of evidence and impaired decision 
making. A summary and discussion of our findings are given in Sec.~5. Possible applications and 
future directions for our studies are presented in Sec.~6.

%
%
%
%
%
%
\section{Background}
\label{Background}

This work merges three distinct areas of study.  As such, a brief review of the
critical elements associated with each area is given.   Specifically, we review
the DM framework along with key aspects of the FAS literature arising in 
neurodegeneration and TBI.  Finally, we review recent work on compromised
spike train propagation due to FAS.  The three areas are used simultaneously
to produce an innovative viewpoint on cognitive deficits.

%
\subsection{Decision making and diffusion-to-boundary models}
%
%

Decision Making (DM) remains an active topic of research and is studied across 
a large variety of fields such as psychology, economics, engineering (and many others), 
but most importantly, in neuroscience.  For an overview of DM theory, see Shadlen and Kiani \cite{Shadlen2016} for 
an up-to-date perspective of the field. In this subsection, we review some key aspects of 
the theoretical and computational models that regard DM as a window on cognition 
\cite{Shadlen2013}. We believe that a proper, quantifiable DM framework is essential 
to evaluate and interpret cognitive deficits arising from traumatic brain injuries and 
neurodegenerative diseases.  See Fig. \ref{DM_scheme} for an illustrative schematics.

In this work, we follow Shadlen et al. \cite{Shadlen2007} and consider \textit{perceptual 
binary decisions} (A vs B) made upon the sequential analysis of evidence. One common 
experimental setup consists of presenting a patch of moving dots to a fixated subject (see 
Panel \ref{DM_scheme}a), where a fraction of the dots move coherently in a given direction. 
At any time, subjects can indicate their choice about the direction of motion of the stimulus. 
The difficulty level of the task is regulated by increasing or decreasing the \textit{coherence level} 
within the stimulus (Panel \ref{DM_scheme}b). This task allows for simultaneous recordings 
of reaction time and perceptual accuracy and is typically modeled by \textit{diffusion-to-boundary}
models \cite{Bogacz2006,Ditterich2006,Ratcliff1998}. As illustrated in Panel \ref{DM_scheme}c, 
\textit{evidence} arrives discretely or continuously in time. In our scheme, 
different signs/colors refer to distinct choices. 
A \textit{decision variable} accumulates evidence in Panel \ref{DM_scheme}d and commits to 
one of the alternatives by reaching a \textit{boundary}.  These tasks are
typically performed by human or monkey subjects, allowing for both psychophysical 
and physiological investigation,  see \cite{Shadlen2013} for a comprehensive overview. 

The diffusion-to-boundary model, although perhaps most commonly used in theory, is only 
one specific manifestation of a DM model.  A variety of other exist which modify this
basic architecture.
Indeed, recent biological experiments in rats and humans, for instance, 
have successfully explored random click tasks to reveal internal properties of decision-making processes
\cite{Brunton2013}, thus providing a modification of the basic DM architecture.
We have chosen to use the basic theoretical framework by
Shadlen et al. \cite{Shadlen2007}, but the others could just as easily have been used
without affecting our basic paradigm.  Specifically, we develop a system with biologically plausible DM capabilities 
coupling an accumulation of evidence model to a neuronal network whose robust population codes
are interpreted as evidence for Choices A or B. We will demonstrate that the presence of
FAS hinders the network functionality in a variety of ways, including by compromising evidence production, leading to 
the anomalous production of evidence and consequently, to accuracy and speed deficits. 
%
%
%
\subsection{FAS in TBI and neurodegenerative diseases}
%

The presence of diffusive FAS is ubiquitous across a host of neurodegenerative
diseases and TBI.  This is well established in the literature that is reviewed below
by disease type.  Although other deleterious effects may be present in a damaged axon, 
the FAS in and of itself can significantly alter spike trains and their information content.  Note
that many of these landmark studies have only appeared in the last 5 years, allowing
for our theoretical developments.

\paragraph{Traumatic Brain Injury:} TBI is one of the major causes of disability 
and mortality worldwide, which in turn, dramatically jeopardizes society in several 
socioeconomic ways~\cite{cdc}. Concussions and other milder forms of TBI are more 
than ever a concern for contact sports and their practitioners~\cite{book_SI}. Soldiers 
are systematically exposed to blast injuries, which led to the recognition of TBI as the 
signature wound of the wars in Iraq and Afghanistan~\cite{Jorge2012}. While many 
survive TBI events, persistent cognitive, psychiatric, and physiological dysfunction often 
follows from the mechanical impact \cite{Lobue2016}. These issues are pushing the 
scientific and medical communities to transform clinical procedures \cite{Yue2013} and 
update our epidemiological understanding of TBI in society \cite{Roozenbeek2013}. 
See Menon et al. ~(\cite{Menon2015}) for a recent report on progress, failures and new 
approaches in TBI research. 

One of the main complications of TBI pathophysiology is that it may affect several
different spatial scales: from cellular/microscopic levels to tissue and/or network 
levels~\cite{Sharp2014}. Despite the insights gained with animal studies 
~\cite{Browne2011,Rubovich2011,Xiong2013} and \textit{in vitro} experiments
~ \cite{Hemphill2011,Hemphill2015,Magdesian2012,Magdesian2016}, there are still 
many open questions and opportunities for translational studies \cite{Hill2016}. In 
particular, there have been only a modest number of theoretical and numerical works 
regarding blast-induced shockwaves \cite{Mau2016}, pathological effects on neuronal 
signaling~\cite{Kolaric2013,Lachance2014} or network dysfunction \cite{Rudy2016}. 
This work aims to contribute with the latter, linking commonly found axonal injuries to 
a network's impaired decision-making capabilities. 

\paragraph{Alzheimer's Disease:} AD is the most common type of dementia -- an umbrella term that 
describes a variety of disorders that arise when neurons die or no longer function normally
 ~\cite{Jorm1998,Thies2013}. Neuronal malfunction ultimately affects memory, behavior and 
the ability to think clearly, advocating for earlier diagnostics of mild cognitive impairments
~ \cite{Petersen2004}. Since aging is the single greatest risk factor for AD \cite{Patterson2015}, most  
public health systems across the world are expected to face huge challenges due to the growing 
elderly population~\cite{Qiu2009}. In fact, someone in America develops AD every 68 
seconds ~\cite{Thies2013}. \\

Recent reports present strong evidence that TBI patients are more likely to develop 
neurodegenerative disorders such as Alzheimer's Disease, Chronic Traumatic Encephalopathy 
and Amyotrophic Lateral Sclerosis \cite{Gupta2016}. The risk is increased for those who, in 
adulthood, sustained a severe head injury \cite{Ikonomovic2004}. See 
\cite{Barnes2014,Johnson2010,Johnson2012,Lobue2016} and references therein for more studies 
linking TBI to AD and other forms of dementia. One commonality between all previously mentioned 
brain disorders (and many others) is the extensive presence of axonal injury, which we will explore 
in details in this work.

\paragraph{Focal Axonal Swellings in TBI:}
Axonal injury (also referred to as diffuse/traumatic axonal injury) is a major outcome of all 
severities of TBI ~\cite{Edlow2016,Hanell2015,Henninger2016,Hill2016}. The longstanding 
assumption that axon loss is an immediate consequence of impacting mechanical forces has been 
supplanted by the understanding that most injured axons undergo secondary progressive changes 
that culminate in Focal Axonal Swellings (FAS) ~\cite{Hemphill2015,Johnson2013,Reeves2012}. 
FAS nomenclature varies across the literature, with varicosities, bulbs, spheroids, torpedoes and 
beadings being common synonyms. In any case, impaired axons provide a clear biophysical marker 
for evaluating cognitive and behavioral deficits induced by TBI~ \cite{Adams2011,Hay2016,Skandsen2010}. 

The development of FAS following TBI is studied both \textit{in vivo}   
\cite{Browne2011,Dikranian2008,Maxwell1997,Wang2011} and \textit{in vitro} experiments
\cite{Chen2009,Fayaz2000,Hellman2010,Hemphill2011,Hemphill2015,Magdesian2012,Morrison2011,Smith1999}
and is tracked whenever possible in human patients 
\cite{Adams2011,Blumbergs1995,Christman1994,Grady1993,Jorge2012,Kinnunen2010,Povlishock2005}.
Many factors can influence the morphological changes in axons ~\cite{Hemphill2015,Johnson2013,Morrison2011}, 
but the most striking finding is that swollen axon diameters can grow dramatically. Tang-Schomer et al. 
\cite{TangSchomer2012,TangSchomer2010}, for instance, reported significant axon diameter increases. 
Such changes are expected to significantly impair spike train propagation and consequently, the information encoded 
in them. 

\paragraph{FAS in AD and other neurodegenerative diseases:} 
Axonal swellings are implicated in a large variety of neuronal disorders such as 
Alzheimer's Disease (AD)~ \cite{Adalbert2009,Daianu2016,Krstic2012,Tsai2004}, 
Multiple Sclerosis~ \cite{Friese2014,Nikic2011,Trapp2008}, 
Parkinson's Disease ~\cite{Tagliaferro2016,Louis2009,Galvin1999}, 
Creutzfeldt-Jakob's Disease~ \cite{Liberski1999},
HIV Dementia~ \cite{Adle1999},
Neuromyelitis Optica~ \cite{Herwerth2016},
Neuropathies~ \cite{Karlsson2016,Lauria2003}, 
and Pelizaeus-Merzbacher Disease \cite{Laukka2016}.
In many cases, FAS arise by the agglomeration of specific proteins over time 
\cite{Coleman2005,Millecamps2013}. In AD, for example, cell stress induces 
accumulation of amyloid precursor proteins in axonal compartments, Overall, one 
of the challenging aspects of research in neurodegenerative diseases and TBI is 
understanding how neuronal pathologies developed at a cellular level will ultimately 
compromise the functionality of an entire network of neurons. 

%
\subsection{FAS and their effects to spike train propagation}
\label{FAS_prop}
%
%

\begin{figure}[t]
\centering
\centering{\includegraphics[width= 0.4\textwidth]{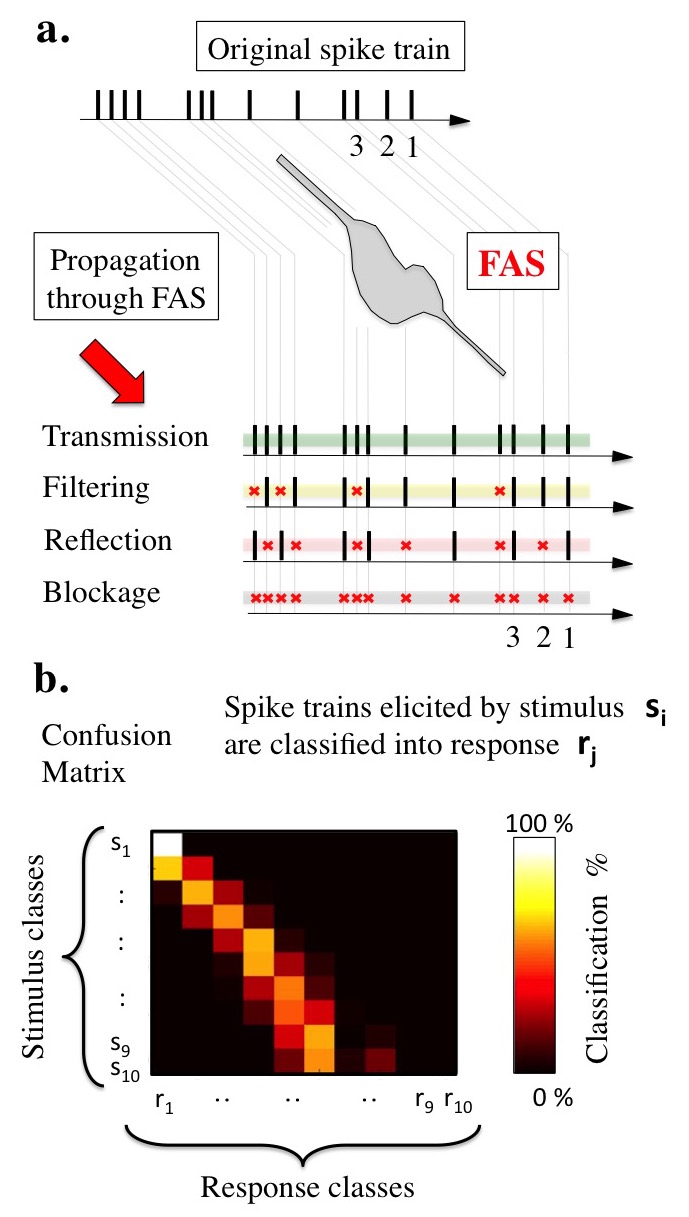}}
\caption{ Schematics for injurious effects of Focal Axonal Swellings (FAS) to spike 
trains according to the theoretical framework of Maia and Kutz \cite{Maia2014_1,Maia2014_2,Maia2015}. 
~\textbf{a.} Spike trains may be distorted in four qualitatively distinct regimes depending on geometrical 
parameters associated to FAS: transmission, filtering, reflection or blockage. See references and text 
for more details. Injuries can then be characterized by a transfer function that returns the effective firing 
rate after spike propagation through the FAS. Such distortions are incorporated to injured neurons in our 
network simulations in conjunction with experimental morphometric results  from Wang et al. \cite{Wang2011}. 
~\textbf{b.} Confusion matrix associated to filtering regime, where firing rate $s_{i}$ may be confused with
rate $r_{j}$ after FAS, adapted from \cite{Maia2014_2}. 
}
\label{spike_scheme}
\end{figure}

In the previous section, we argued that FAS are almost universal in neurodegenerative 
diseases, concussions, and TBI. However, as recently pointed by Hill et al. ~\cite{Hill2016}, 
one outstanding question remains unanswered: what is the functional significance of axonal 
varicosities? We believe that axonal swellings are responsible for, among other things, hindering the information 
transmitted from injured neurons to others. 
It is broadly accepted that (healthy) neurons excite/inhibit their neighbors proportionally 
to their firing-rate and that this collective activity ultimately produces some form
of high-level network functionality (decision-making capability, memory retrieval, learning, etc). 
Recent theoretical and computational works demonstrate that there is an injurious
interplay between electrophysiological dynamics and FAS.  We will review some of these 
results and present ways to jeopardize neuronal firing-rate capabilities in a biophysically 
plausible manner. By modifying the internal dynamics of injured neurons, we obtain a large 
variety of interpretable network deficits.

\paragraph{Distinct propagation regimes:} Maia et al.~ \cite{Maia2014_1} investigated 
cable equations with varying axonal diameters to identify critical regions for spike propagation 
in axon segments. They characterized FAS morphologies by specifying key geometrical parameters 
in a prototypical axonal enlargement model; the total diameter increase and how abruptly it occurs. 
These parameters determine the propagation regime that spike trains will undertake when they reach 
the FAS (See the schematics in Fig. \ref{spike_scheme}a). Smaller swellings may still allow for intact 
transmissions of spikes although the filtering regime (with spike shifts and deletions) is more common. 
A spike can also split into two, one traveling forward and the other traveling backward. In this case, 
when the backward propagating pulse collides with the next spike in the original spike train, both spikes 
are deleted. As a consequence, a neuron's firing rate in a reflection regime is effectively halved. 
For more dramatic morphologies, the spike train can be completely blocked, setting the neuron's firing 
rate to zero. Axonal injury can thus be modeled by a (parameter-dependent) transfer function that returns 
the effective firing rate after spike propagation through the FAS. The filtering regime was explored in details 
in \cite{Maia2014_2}, where a pile-up collision mechanism deletes/filters one (out of two) adjacent spikes. 
Thus, there is a probability that a spike train with firing rate $s_{i}$ will be confused as having firing rate 
$r_{j}$ instead of $r_{i}$. The confusion matrix in Fig. \ref{spike_scheme}b summarizes all distinct scenarios. 
Notice that higher firing rates are more affected, distorting the diagonal structure of the matrix. \\

The limitations of this methodology and its applicability to FAS diagnostics are explored in~\cite{Maia2015}. For instance, the theoretical results are valid only for unmyelinated axons
and should not be inadvertently extrapolated to other biophysical settings. From a pathological 
perspective~\cite{Reeves2012}, unmyelinated axons are more vulnerable to TBI and comprise the numerical 
majority of cerebral axons. 

%
%
%
%
%
%
\section{Materials and Methods}
\label{Methods}

With the background theory in hand, we can develop an integrated theoretical approach to understanding
the effects of neurodegeneration and TBI in neuronal networks responsible for decision making.  The specifics
of our theory are now given along with how the FAS statistics are used to calibrate our model.

%
\begin{figure*}
\centering{\includegraphics[width=0.7\textwidth]{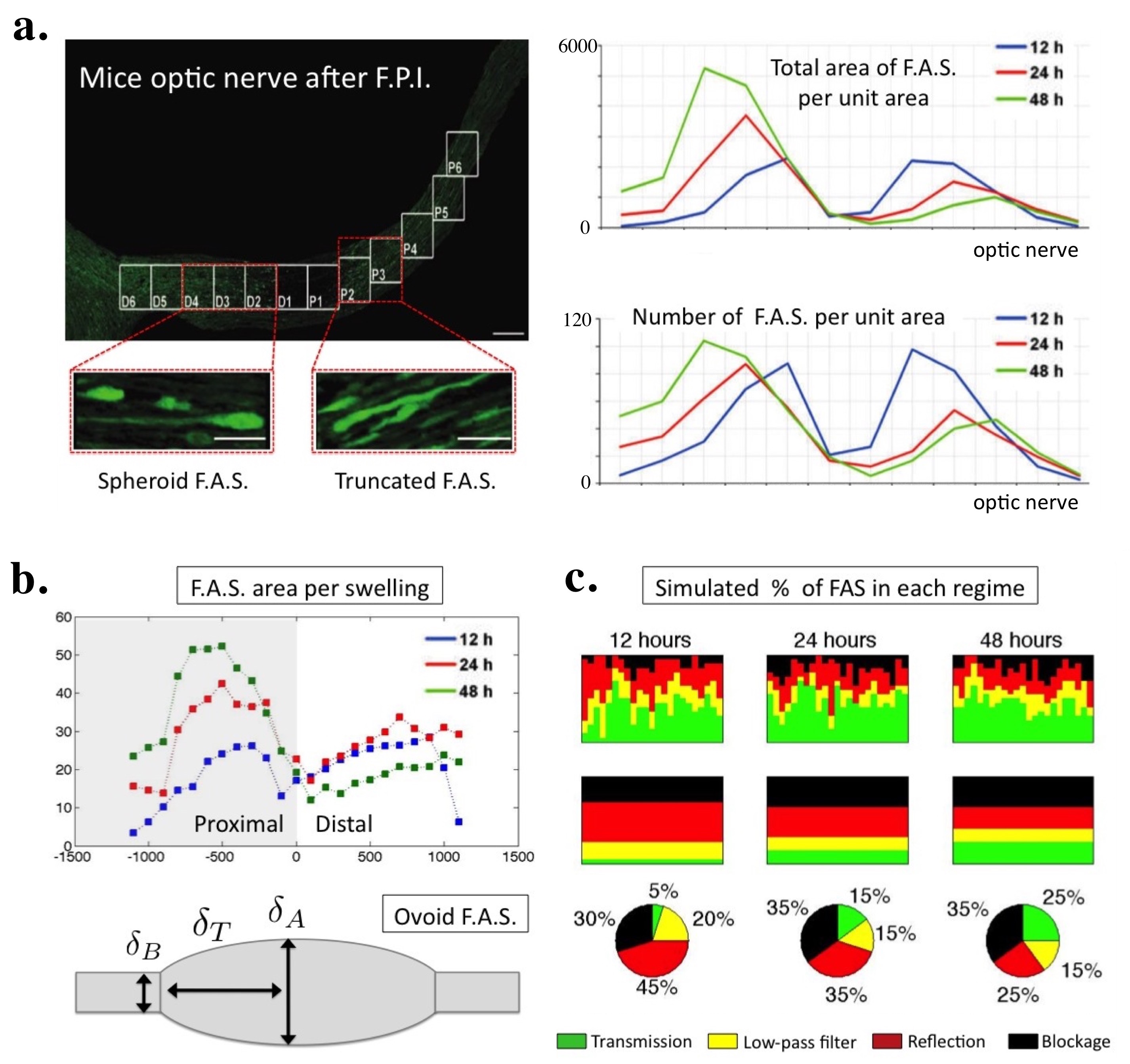}}
\caption{Inferring Focal Axonal Swellings (FAS) parameters from the experimental 
work of Wang et al. (\cite{Wang2011}). 
~\textbf{a.} 
In their setup, they traumatically 
induce axonal injury in mice by targeting its optical nerve. Fluorescence analysis
and confocal microscopy reported \textit{qualitatively} different shapes of FAS: 
spheroids (ellipses) at the distal segment and truncated shapes at the 
proximal segment. The total area and number of FAS varied
as the injury progressed in time (12h, 24h and 48h).
~\textbf{b.} 
We generate a family of ovoids having area within 
one standard deviation from the distributions of FAS area 
per swelling and compatible with the experimental distribution. 
For each generated FAS, we extract the parameters 
$[\delta_{B}, \delta_{T}, \delta{A}]$, necessary to estimate 
the propagation regime $\eta$ (following \cite{Maia2015,Maia2014_2,Maia2014_1}). 
~\textbf{c.}
We generate 12 FAS (column) for each injured axon (row) and order them 
from worst to best-case scenario (upper ``flags''). We assume that the worse FAS 
within an injured axon dominates the others, and classify the entire 
axon within that category (intermediary ``flags''). The procedure yields reasonable
estimates of the \% of FAS in transmission, filtering, reflection and blockage regimes  
(bottom pie-charts of impairments for an injured neuronal population).  
}
\label{FAS_and_PIES}       
\end{figure*}
%
%
%
\subsection{Inferring FAS parameters from experimental data}
\label{infer_fas_param}
%
%
%
The morphological features of FAS implicated in TBI 
and neurodegenerative diseases are heterogeneous. They can also 
jeopardize individual and collective neuronal dynamics in different 
ways. While injured networks are expected to perform worse, the 
main challenge is to introduce pathologies to the computational model 
in a biophysically compatible way. In this regard, we will infer the 
distribution of FAS and their corresponding functional deficits from 
a set of experimental results.  

Wang et al. (\cite{Wang2011}) traumatically damage the optic nerve of
adult rats with a central fluid percussion injury. They sacrifice mice at 
different times after the TBI impact (12h, 24h, and 48h respectively) and 
remove the optic tract from the skull for fluorescence analysis and 
confocal microscopy. Figure \ref{FAS_and_PIES}a shows this relatively 
organized and aligned bundle of axons. Two qualitatively different shapes of 
swellings  develop: spheroids (ellipses) at the distal segment and truncated 
forms at the proximal segment. The most relevant aspect of their study 
for us is the assessment of the total area and the number of FAS 
(per unit area). See the right plots in Fig. \ref{FAS_and_PIES}a. 
Swelling distributions differ along the spatial segments of the bundle 
in shape, size and number. Moreover, they evolve dynamically 
as the injury progress in time.  

In Fig. \ref{FAS_and_PIES}b, we estimate the average area per swelling. 
Since the authors describe FAS in the distal segment as elliptical shapes
(spheroids), we can generate a family of ovoids having an area 
within one standard deviation of the measurements. For each generated 
FAS, we extract the parameters $[\delta_{B}, \delta_{T}, \delta{A}]$, necessary 
to estimate the spike propagation regime (see section \ref{FAS_prop} or 
\cite{Maia2014_1,Maia2014_2,Maia2015}). 
Wang et al. \cite{Wang2011} divide the optic nerve into 12 spatial grids, and we 
generate one swelling per grid for each injured axon. The upper flag-charts in 
Fig. \ref{FAS_and_PIES}b show 20 injured axons (rows) with 12 FAS each (columns). 
Each FAS has a functional deficit (transmission, filtering, reflection or blockage) 
according to its geometrical parameters. Finally, we assume that the worst swelling 
within an axon dominates and classifies the entire axon within that category 
(intermediate flag-charts). This lead to the pie-charts indicating the fraction of axons 
in each propagation regime. 

There are several drawbacks with this methodology, but we believe that our distributions 
are biophysically reasonable and compatible with existing available data. Better results could be 
obtained if Wang et al. \cite{Wang2011} used the recently developed diagnostic (computational) tool of 
Maia and Kutz \cite{Maia2015}. 
Still, the results are interesting, because although the fraction of FAS in the transmission 
regime increases from 12h to 48h, the fraction of swellings in the blockage regime increases 
as well. Thus, it is not obvious if the system is ultimately recovering or losing functionality
(see Fig. \ref{FAS_and_PIES}b). 
%
%
\subsection{Network model and governing equations}
%
%
%
The modeling of neuronal networks is one of the most vibrant fields of 
computational neuroscience; there are numerous models throughout the 
literature, with varying levels of complexity, architectural configurations, and 
biological functionality~\cite{Network_Review,Dayan2001,Gerstner2014}. 
In this work we build a system with decision-making capabilities from 
a network of synaptically coupled firing rate units \cite{Dayan2001}.
This setup allows the addition of injurious effects of FAS to neuronal dynamics 
in an intuitive and simple manner while retaining key aspects of the works 
of Maia and Kutz \cite{Maia2014_2,Maia2014_1}. 

More broadly, high-dimensional neuronal networks are ubiquitous and characterized by a large connectivity graph whose structure determines how the system operates as a whole~\cite{Watts:1998db,Park2013science}.  Typically the connectivity is so complex that the functionality, control and robustness of the network of interest is impossible to characterize using currently available methods.  Moreover, with few exceptions, underlying nonlinearities impair our ability to construct analytically tractable solutions, forcing one to rely on experiments and/or modern high-performance computing to study a given system. However, advances over the past decade have revealed a critical, and seemingly ubiquitous, observation:  that meaningful input/output of signals in high-dimensional networks are encoded in low-dimensional patterns of dynamic activity~\cite{jones07,rabinovich01,rabinovich08}.

\begin{figure}[th]
\centering
\centering{\includegraphics[width= 0.4\textwidth]{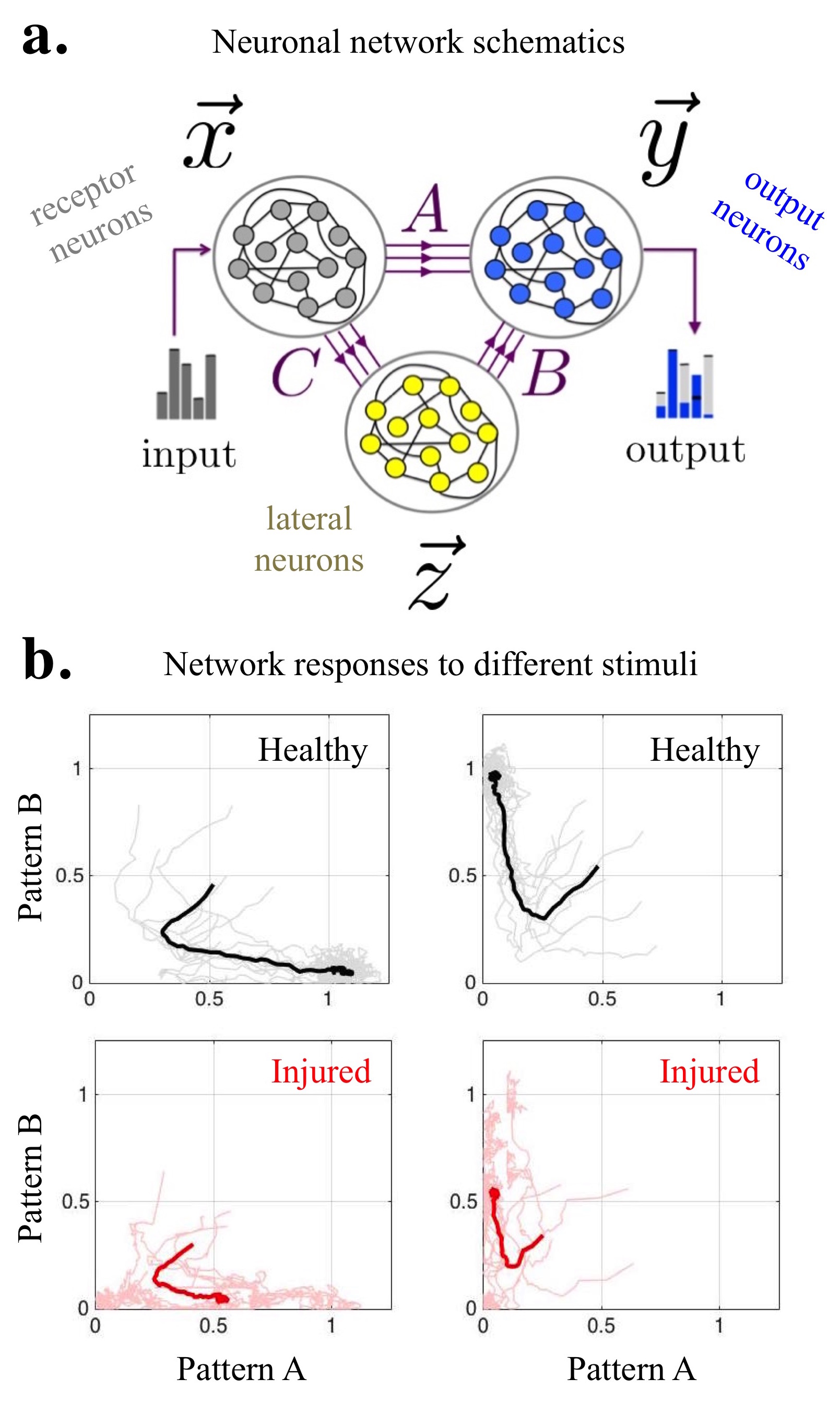}}
\caption{Neuronal network schematics and its dynamical responses to different stimuli. 
~\textbf{a.} We consider three distinct subpopulations: receptor neurons
($\vec{x}$), projection neurons ($\vec{y} $) and lateral 
neurons ($\vec{z}$). Receptor neurons receive direct input from external 
stimuli while projection neurons act as readout/output sources. The matrices 
$A,B$ and $C$ represent the strength of the connections between the neurons from 
different populations. See Equation (\ref{NetEqs1})--(\ref{NetEqs3}) for the 
dynamics of the model and Table \ref{NetSymbols} for a full
list of parameters and variables.
~\textbf{b.} Collective responses of projection neurons produce stereotypical patterns in a 
proper low-dimensional PCA space: a healthy network (in black) consistently drives the responses
to corresponding fixed points; left plot for Stim. A, and right plot for Stim. B. 
The output patterns of an injured network (in red), however, may vary significantly and 
a significant number of trajectories might not reach the expected fixed points. 
}
\label{Net_&_Traj}
\end{figure}

We avoided committing to a specific network architecture, keeping only
three broad classes of neuronal subpopulations (see Fig. \ref{Net_&_Traj}a):
input neurons $\vec{x}$ that receive different stimuli from the external environment, 
output neurons $\vec{y}$, and local/lateral interneurons $\vec{z}$. 
The dynamical equations of the system are given by: 
\begin{eqnarray}
\label{NetEqs1}
\text{d} \vec{x} & = & - \vec{x} ~\text{dt} + \bold{S}(t) \text{dt} + \mu \text{dW} \\
\text{d} \vec{y} & = & -  \beta \vec{y} ~\text{dt} + \big[ A ~ \vec{x} - B~\vec{z}\big]^{+} \text{dt}  + \mu \text{dW}  \\
\text{d} \vec{z} & = & - \gamma \vec{z} ~\text{dt} + \big[ C~\vec{x} - E~\vec{z}\big]^{+} \text{dt}  + \mu \text{dW}
\label{NetEqs3}
\end{eqnarray}

\noindent The operator $[ ~ ]^{+}$ is a standard linear threshold function (\cite{Dayan2001}) that
rectifies each component of the vector, i.e. 
\begin{equation}
\big[~ \xi ~\big]^{+} = 
\left\{
\begin{array}{l l l}
    0 & \quad \text{if $~\xi \leq 0$,} \\
    \xi & \quad \text{if $~0 < \xi < 1$,} \\
    1 & \quad \text{if $~\xi \geq1$ .}
  \end{array}
  \right.\
\label{threshold}
\end{equation}
\noindent The connectivity matrices $A,B,C$ and $E$ account for the neuronal coupling terms.
See Table \ref{NetSymbols} for a list of all parameters used in the model. 
The network was calibrated (as in \cite{Shlizerman2014}) to produce distinguishable 
responses to meaningful external stimuli and match observed input/output patterns. 
The input neurons $\vec{x}$ lock at a faster timescale onto the driving stimulus $\bold{S}(t)$, 
which in turn, excite or inhibit the $\vec{z}$ subpopulation so that $\vec{y}$ may 
produce a stable output associated to it. 
%
%
\begin{table}
\centering
\caption{Variables and parameters for neuronal network model}
\scriptsize
\begin{tabular}{ c | c | l }
Symbol & Description & Remarks
 \\ \hline
 & & \\
$\vec{x}$ &   input/receptor neurons & $N$ firing-rate units \\ 
$\vec{y}$ &   output/projection neurons & $N$ firing-rate units \\ 
$\vec{z}$ &   lateral/local interneurons  & $N$ firing-rate units \\ 
& & \\
$1$ & time scale for dynamics of $\vec{x}$  & fast, reference time scale\\
$\beta^{-1}$ & time scale for dynamics of $\vec{y}$ & slower time scale \\
$\gamma^{-1}$ & time scale for dynamics of $\vec{z}$ & slower time scale \\
& & \\
$A$ & connections between  $\vec{x}$ and $\vec{y}$  & assumed to be known \\
$B$ & connections between  $\vec{z}$ and $\vec{y}$ & calibrated to match i/o patterns\\
$C$ & connections between  $\vec{x}$ and $\vec{z}$ & assumed to be known \\
$E$ & connections between  $\vec{z}$ and $\vec{z}$ & randomly connected \\
& & \\
$\bold{S}(t)$ & external stimulus & usually $\bold{S_{A}}$ or  $\bold{S_{B}}$ \\
$\mu$ & noise intensity & assumed Brownian ($dW$) \\
$[~~ ]^{+}$ & linear threshold function & see Equation (\ref{threshold}) 
\end{tabular}
\label{NetSymbols}
\end{table}

To simulate our equations numerically, we modified the MATLAB implementation 
of the Euler-Maruyama scheme from D. J. Higham \cite{Higham2001} to solve 
the stochastic differential Eqs.~(\ref{NetEqs1}) - (\ref{NetEqs3}). The 
$\text{dW}$ term refers to a standard Brownian/Wiener process, and the noise 
intensity level $\mu$ was calibrated to model natural stochastic fluctuations in the system. 

%
%
%
%
%
\subsection{Network responses in proper low dimensional spaces}
%
%
The output neurons in the network produce high-dimensional dynamical 
responses $\vec{y}(t) \in \mathbb{R}^{N}$, although they do not always have  
biological relevance. For example, in the absence of external stimulus, 
$\bold{S}(t) = 0$, the network's activity consists of simple stochastic fluctuations
around the resting state of the system. Thus, we will focus on the outputs of a pre-defined
set of external stimuli ($\bold{S_{1}}, \bold{S_{2}},\ldots $) whose meaning should
vary with context.  For instance, on a network representing the antennal lobe of the \textit{Manduca sexta} moth, the output codes  represent the response to different stimulus odors used in 
flower discrimination (\cite{Riffell2014,Shlizerman2014}).   In the context of
cognition, the external stimuli would be, for instance, associated with a recognition task (visual, auditory, tactile, etc).

The corresponding network responses form a library
$$
L = \begin{bmatrix}
      \vdots  &  &      \vdots      \\
       \vec{y}^{P}_{1}& \cdots & \vec{y}^{P}_{l} \\
       \vdots & & \vdots
     \end{bmatrix}_{N \times l }
$$
in which each element (or pattern) encodes a specific stimulus:
$ \vec{y}^{P}_{1}$ encodes $\bold{S_{1}}$, 
$ \vec{y}^{P}_{2}$ encodes $\bold{S_{2}}$, and so on.
With a proper transformation and change of basis,
one can rewrite the library as a set of orthonormal vectors 
$$
O = \begin{bmatrix}
      \vdots  &  &      \vdots      \\
       \vec{o}^{P}_{1}& \cdots & \vec{o}^{P}_{l} \\
       \vdots & & \vdots
     \end{bmatrix}_{N \times l .}   
$$
See \cite{Shlizerman2014} for mathematical and technical details.
This allow us to write the dynamics of the output neurons with the 
low-rank decomposition
\begin{equation}
\vec{y}(t) = p_{1}(t) \vec{o}^{P}_{1} + \cdots + p_{l}(t)\vec{o}^{P}_{l} 
+ r(t) \vec{o}^{R}.
\label{eq_lowdim}
\end{equation}
What completes the decomposition is the \textit{remainder} 
vector $\vec{o}^{R}$ onto which all meaningless features (non-library elements)
are projected. We can now easily interpret network activity over time by tracking the projections $ p_{1}(t), p_{2}(t),\ldots, p_{l}(t)$
and $r(t)$.

\paragraph{Binary case:} In this work, we limit our library to 
two distinct patterns ($A$ and $B$) and use them to build a system 
with binary decision-making capabilities.  This restriction can easily be lifted, but
for the purposes of illustrating compromised decision making, a binary task will
serve to illustrate our theory best.   In this case, Eq.~(\ref{eq_lowdim}) simplifies to 
$$
\vec{y}(t) = p_{A}(t) \vec{o}^{A} + p_{B}(t) \vec{o}^{B} + r(t) \vec{o}^{R}.
$$
In Figure \ref{Net_&_Traj}B, we randomly initiate a healthy network and stimulate it 
with $\bold{S}(t) = \bold{S_{A}}$ (left plots) or $\bold{S}(t) = \bold{S_{B}}$ (right plots) and depict several 
trajectories (in gray) projected onto the PCA subspace generated by $\{ \vec{o}^{A}, \vec{o}^{B}\}$. 
Without noise (see \cite{Shlizerman2014}), the trajectories would reach the fixed points $(1,0)$ or $(0,1)$. 
Instead, they randomly fluctuate/gravitate around these values while the stimulus persists. 
The outcome for injured networks can vary significantly (see trajectories in red), 
depending on the injury levels, the subset of targeted neurons and FAS morphological 
distributions. These discrepancies will ultimately hinder the network's production of evidence 
and impair its decision-making abilities. 
%
%
%
%
%
%
\section{Results}
\label{Results}

\subsection{Compromised network responses}
%
%
%
\begin{figure}[t]
\centering
\includegraphics[width=0.5\textwidth]{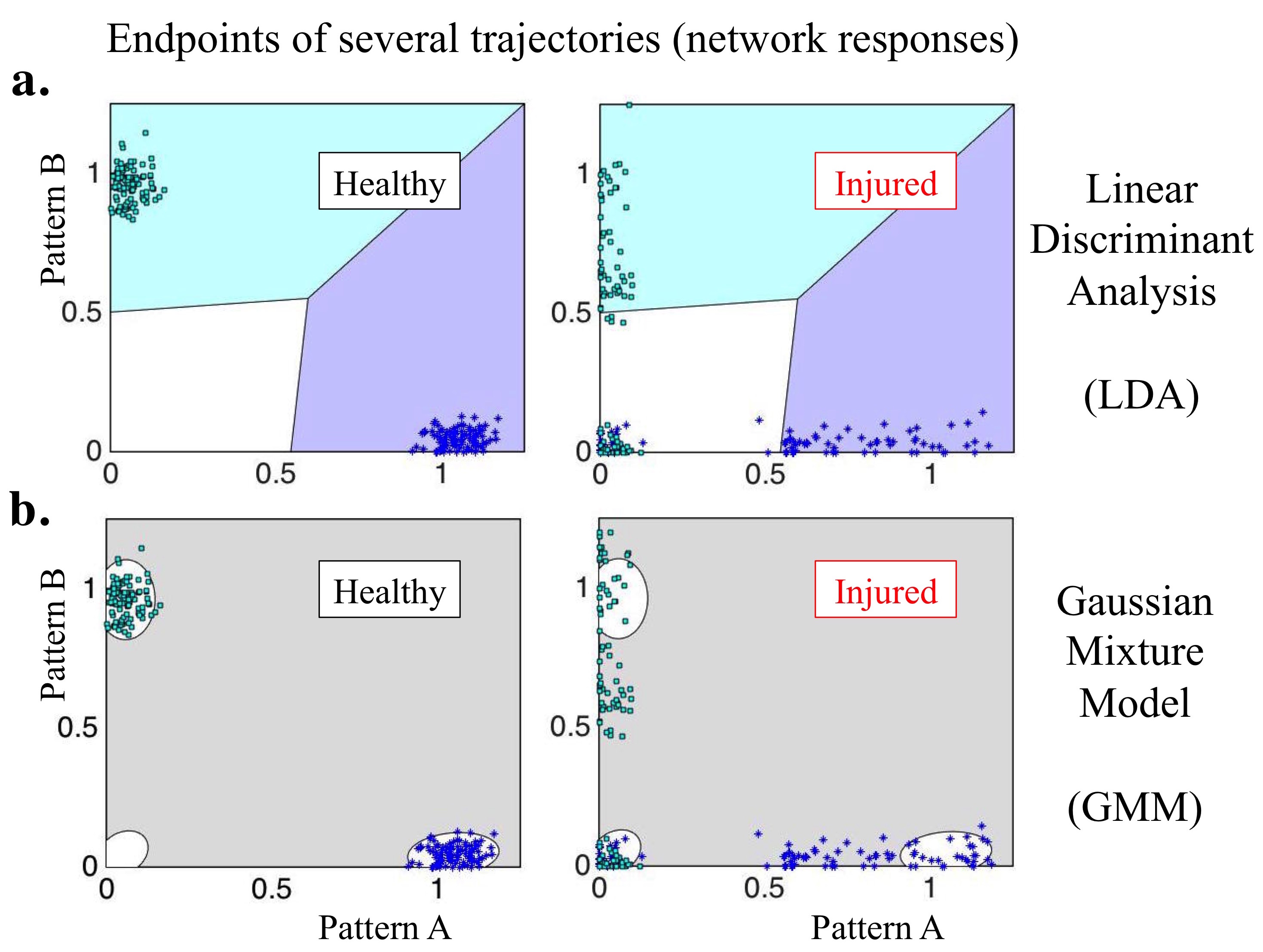}
\caption{Endpoints for several trajectories. Network responses triggered by stimulus A/B
are depicted as dark/light blue dots respectively. Notice that the endpoints of a healthy 
network (left plots) pile-up/gravitate around fixed points for the corresponding stimuli. 
The responses of an injured network (right plots) end significantly more scattered. 
We use two different criteria to discriminate between responses and define evidence-producing
regions. 
~\textbf{a.} Linear Discriminant Analysis (LDA) finds the optimal linear classifier between 
distinct response patterns.
~\textbf{b.} Gaussian Mixture Model (GMM) yields the best 3-gaussian fit to 
classify the distinct responses, with 95\% confidence intervals (ellipses) depicted in white.   
In both cases, several responses of the injured network are misclassified.}
\label{LDA_and_GMM}       
\end{figure}
%
%
%
We showed that a healthy network responds to meaningful stimulus 
with a trajectory (projected in a proper low dimensional space) that 
ends near a fixed point of the system. This behavior can change 
significantly with injury; see Figures \ref{Net_&_Traj}b and \ref{LDA_and_GMM} 
for illustrative examples. In these cases, we randomly added FAS to 30\% of 
the neurons in the network, with 10\% leading to the filtering regime, 10\% 
to reflection and 10\% to the blockage regime. Note that the trajectories do 
not form clear clustering regions around the fixed points anymore. Instead, 
many trajectories end near the origin, which means that no stable response 
pattern (library element) was produced. We use two different criteria to define 
these clustering regions and quantify the number hindered/misclassified responses.

\paragraph{Linear Discriminant Analysis:} LDA is a method used in statistics,  
pattern recognition and machine learning to find a linear combination of features  
which separates two or more classes of objects or events. In our case, we use LDA 
to separate the network healthy/typical responses in three classes:  A, B, and Resting 
State (when no stimulus is presented).
We generate a training set for which the correct classification is known and find a good 
predictor for any sample with the same distribution (not from the training set so as
to cross-validate the findings). 
This results in hyperplanes (lines in the 2D case)  that yield maximum separation between 
distinct classes. See Figure \ref{LDA_and_GMM}a for an example of LDA classification; the 
dark blue region corresponds to $\bold{R_{A}}$, the light blue region to $\bold{R_{B}}$ and 
the white region to the resting state.  
The left panel of Fig. \ref{LDA_and_GMM}a shows that many injured responses can 
cross the original separating lines, with most of them falling in the white region. The  
misclassification of a response $\bold{R}$ to its corresponding stimulus $\bold{S}$
can be quantified by a pair confusion matrices (before/after injury): 

%
%
\begin{eqnarray}
\centering
&& \begin{tabular}{c | c c c}
Pre injury ~& $\bold{R_{A}}$& $\bold{R_{B}}$& Rest \\
\hline
$\bold{S_{A}}$ & \cellcolor[gray]{0.8}  100\% & 0 &  0  \\
$\bold{S_{B}}$ & 0 & \cellcolor[gray]{0.8}  100\% & 0  \\
zero & 0 & 0 &  \cellcolor[gray]{0.8}  100\% \\
\end{tabular}  
\label{conf_1} \\
&& \begin{tabular}{c}
$ ~~~~~~~~~~~~~~~~~~~~~  \downarrow ~~ \textbf{\textit{Injury}} ~~ \downarrow $\\
\end{tabular}
\nonumber 
\\
&& \begin{tabular}{c | c c c}
Post injury & $\bold{R_{A}}$& $\bold{R_{B}}$& Rest \\
\hline
$\bold{S_{A}}$ & \cellcolor[gray]{0.8}  66\% & 0 &  34\%  \\
$\bold{S_{B}}$ & 0 & \cellcolor[gray]{0.8}  62\% & 38\%  \\
zero & 0 & 0 &  \cellcolor[gray]{0.8}  100\% \\
\end{tabular}  
\label{conf_2} 
\end{eqnarray}
%
%
The confusion matrix associated with the injured network is significantly 
less diagonal than the one associated to the healthy network. This is 
a common feature across all injury distributions and will always lead to 
a relative loss of transmitted information~\cite{Maia2014_2}. 

\paragraph{Gaussian Mixture Model:}
GMM refers to a probability density function represented
as a weighted sum of Gaussian component densities and 
commonly used for data clustering. In our case, we use GMM to 
cluster the network responses in the same three distinct classes as before. 
The method uses an iterative algorithm and yields the best 3-Gaussian 
fit to our data. Computations were performed using the \texttt{gmdistribution} 
function from MATLAB's Statistics Toolbox. Figure~\ref{LDA_and_GMM}b 
shows that after injury, many trajectories do not end within the original ellipses
(95\% confidence interval for each gaussian component). The analogous confusion 
matrices (before and after injury), for the same example, are now given by: 
%
%
%
\begin{eqnarray}
\centering
&& \begin{tabular}{c | c c c}
Pre injury ~& $\bold{R_{A}}$& $\bold{R_{B}}$& Rest \\
\hline
$\bold{S_{A}}$ & \cellcolor[gray]{0.8}  95\% & 0 &  5\%  \\
$\bold{S_{B}}$ & 0 & \cellcolor[gray]{0.8}  96\% & 4\%  \\
zero & 0 & 0 &  \cellcolor[gray]{0.8}  100\% \\
\end{tabular}  
\label{conf_3} 
\\
&& \begin{tabular}{c}
$ ~~~~~~~~~~~~~~~~~~~~~  \downarrow ~~ \textbf{\textit{Injury}} ~~ \downarrow $\\
\end{tabular}
\nonumber 
\\
&& \begin{tabular}{c | c c c}
Post injury & $\bold{R_{A}}$& $\bold{R_{B}}$& Rest \\
\hline
$\bold{S_{A}}$ & \cellcolor[gray]{0.8}  25\% & 0 &  75\%  \\
$\bold{S_{B}}$ & 0 & \cellcolor[gray]{0.8}  16\% &  84\%  \\
zero & 0 & 0 &  \cellcolor[gray]{0.8}  100\% \\
\end{tabular} 
\label{conf_4}  
\end{eqnarray}
%
%
%
The GMM criterium is stricter than the LDA criteria and significantly increases 
misclassification after injury. The relative loss of Transmitted Information (TI) 
in this case is equal to 83.96\%, as opposed to a 53\% loss given by the LDA 
criterium. We will discuss the loss of TI in more details in the following section.  
%

%
\subsection{Loss of transmitted information}
%
%
%
\begin{figure}[t]
\centering
\includegraphics[width=0.45\textwidth]{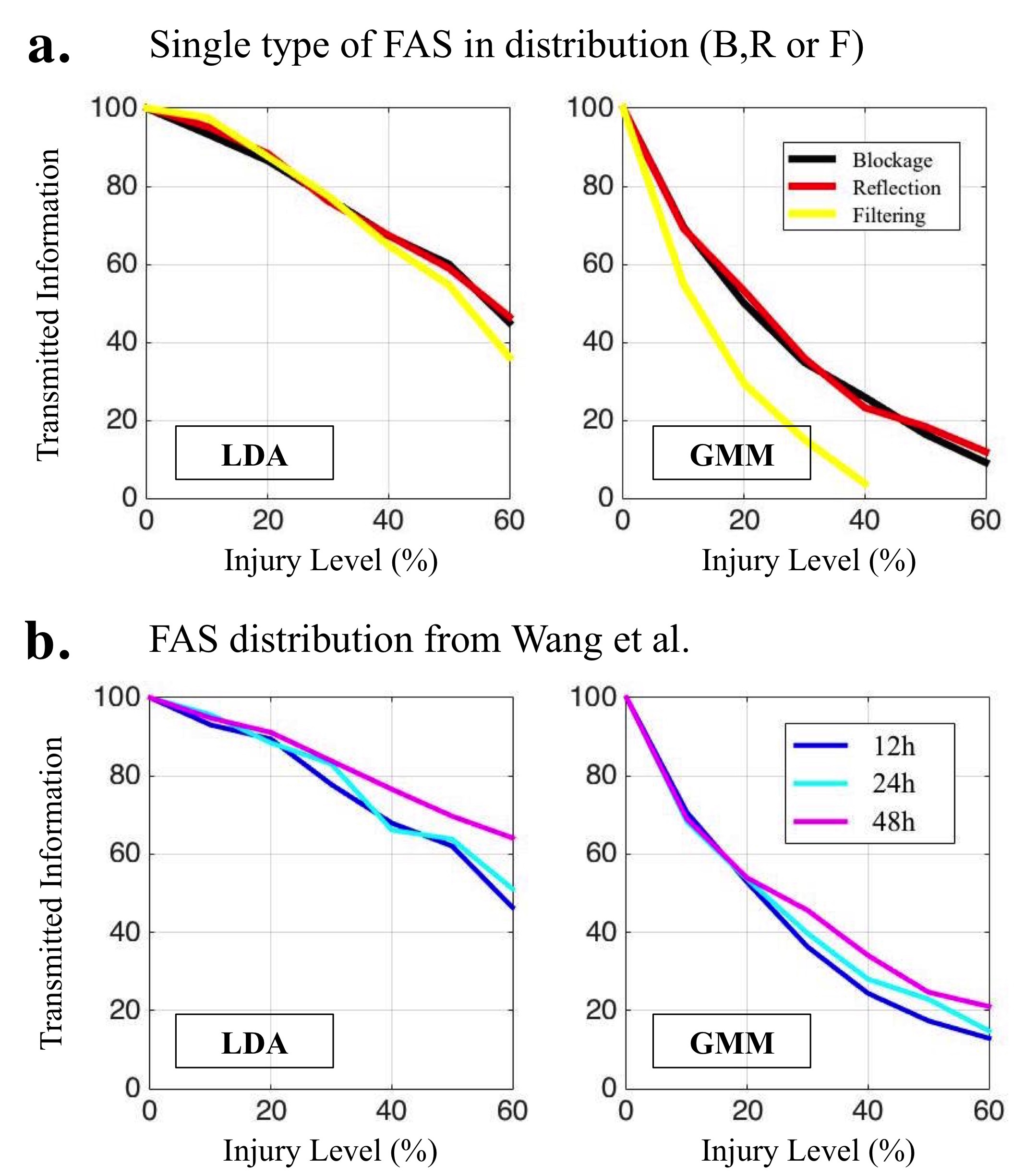}
\caption{Transmitted Information (\%) as a function of injured 
neurons subpopulation (\%). ~\textbf{a.} Here we assume that all 
FAS have the same dysfunctional type of regime: filtering (yellow), 
reflection (red) and blockage (black). 
~\textbf{b.} Here we assume that the fraction of FAS in different 
dysfunctional types follow the pie charts in Figure \ref{FAS_and_PIES}c. 
They correspond to the three different heterogenous distributions of 
FAS (for 12h, 24h and 48h) estimated from the work of Wang et al. 
(\cite{Wang2011}). 
Left  panels use Linear Discriminant Analysis (LDA) as a classification 
criteria while the right panels use a Gaussian Mixture Model (GMM). 
All panels show results for a 60-neuron network 
(with 20 neurons for each subpopulation).}
\label{Loss_Info}       
\end{figure}
%
%
%
The Transmitted Information (TI) in a system quantifies how much its responses 
(to different stimulus) are non-random. It is a quantity calculated from confusion matrices such as those 
from equations (\ref{conf_1})--(\ref{conf_4}), and we refer to  \cite{Maia2014_2,Victor1997} for a more
detailed account of the topic. The matrix in (\ref{conf_1}) has perfect diagonal structure and 
exemplifies maximum TI. Since there are only three stimulus and response classes, we 
have in this case $H = \log_{2} 3 \approx 1.58$. Conversely, the matrix in (\ref{conf_4}) transmits
very little information. 

Intuitively, TI penalizes imperfect confusion matrices proportionally to the magnitude 
of their off-diagonal terms. The misclassification terms grow whenever the system 
responds to stimulus $\bold{S_{i}}$ with a signal that is classified as from class $\bold{R_{j}}$ 
(with $i \neq j$). For both classification criteria -- LDA and GMM -- the loss in TI 
after injury provides a natural metric for loss of cognitive functionality. Maia and
 Kutz~ \cite{Maia2014_2} provide TI decay rates for individual neurons injured with mild 
FAS. In this section, we provide analogous results, but at a collective network level and 
for different FAS distributions.  

\paragraph{Homogeneous FAS distributions:} As discussed in sections \ref{Background}--\ref{Methods}, 
FAS following injury have a large variety of shapes and sizes, and consequently affect neuronal dynamics 
in different ways. As a first example, however, we consider homogeneous FAS distributions throughout the
network, i.e., when all injured neurons have the same dysfunctional regime (filtering, reflection or blockage 
of spike trains). 
Figure \ref{Loss_Info}a shows the decay of TI as a function of injured neurons (\%) for all three regimes. 
For the GMM criterium, a homogeneous distributions of FAS in the filtering regime is more harmful than one in 
the blockage or in the reflection regimes. This result is surprising, since axons subject to the highest amount of 
damage typically exhibit FAS in the blockage regime (\cite{Maia2014_1}). For the LDA criterium (left plot), 
the type of propagation regime does not seem to influence the loss of TI except for larger fractions of injured 
neurons. There again, the filtering regime is slightly worse than the others. 

\paragraph{FAS distributions from Wang et al. \cite{Wang2011}:}
In Figure \ref{Loss_Info}b, we plot TI as a function of damaged neurons 
for the three heterogeneous distributions of FAS from Fig. \ref{FAS_and_PIES}c (pie charts).  
They correspond to FAS development up to 12h, 24h, and 48h after the TBI impact. 
For a small percentage of injured neurons, different heterogeneities did not significantly alter the decay of TI 
(for both LDA and GMM criteria). For larger fractions, however, the results 
show a clear improvement as the injury progressed in time, suggesting 
that recovering mechanisms might have played a role. 

As a consistency check, we note that the heterogeneous distributions yield better
results than the worst homogeneous distribution (filtering alone). Our heterogeneous 
FAS functional distributions were derived from indirect calculations (see section \ref{infer_fas_param}), 
and  we hope that our results incentivize experimentalists to report more geometrical parameters 
for individual swellings. Nevertheless, we were able to illustrate novel methodologies to quantify 
the loss of functionality in injured networks in a biophysically plausible way.

%
\subsection{Anomalous production of evidence}
%
%
In what follows, we model a system with biologically plausible decision-making capabilities 
using the neuronal input/output network of the last section as its core generator of evidence. The network 
responds to incoherent external stimuli producing mixed pieces of evidence, which in turn, are coupled 
to an accumulation model that ultimately decides in favor of one choice (A vs B). We also 
show how the addition of FAS injuries to the network leads to a large variety of anomalies 
regarding production of evidence. 

\paragraph{Mixed stimuli and coherence levels:}  So far, we only applied constant inputs to 
the network; $\bold{S} (t) = \bold{S_{A}}$ or $\bold{S} (t) = \bold{S_{B}}$. 
On the other hand, organisms are faced with many decisions where the evidence 
is noisy, mixed, or arrives intermittently over time \cite{Shadlen2007}. To account 
for more realistic scenarios, we will consider stimulus of the form 
\begin{eqnarray}
\bold{\tilde{S}} (t) = \sum_{j} I_{A}(t_{j}) + \sum_{k} I_{B}(t_{k}), ~~ \text{where} \label{train_stim}
\\
 I_{A,B }(t_{i}) = \left\{
  \begin{array}{l l}
    \bold{S_{A,B}} & \quad \text{if $t  \in  [t_{i}, t_{i} + \tau]$, }\\
    0 & \quad \text{otherwise.}
  \label{train_stim2}
  \end{array} \right. 
 \end{eqnarray}
In simpler terms, the external input $\bold{\tilde{S}}$ is a mixed train of $\bold{S_{A}}$ 
and $\bold{S_{B}}$ pulses, each with duration $\tau$ and starting times  $t_{j}$ and $t_{k}$ respectively.
We will refer to it as a \textit{Poisson} mixed train of stimulus if the pulse initiation times 
$t_{j}$ and $t_{k}$ are generated by their (own, independent) Poisson process of intensity 
$\lambda_{A}$ and $\lambda_{B}$ respectively. It is relatively straightforward to adjust 
these intensities to achieve the desired coherence level, with $ | \lambda_{A} - \lambda_{B}| \approx 0$ \
denoting incoherent trains and  $ | \lambda_{A} - \lambda_{B}| \gg 1$ denoting coherent ones. 
Such input stimulus is consistent with recent, biophysically relevant DM experiments~\cite{Brunton2013}.

\paragraph{Healthy production of evidence:} The network's response to a 
Poisson mixed train of stimulus can be thought
as a series of simpler trajectories alternating between the vicinities of the fixed points 
$A$ and $B$.  We interpret these vicinities as Evidence-Producing Regions (EPR)  
for choices A and B, although their precise definition may vary according to the clustering 
classification criteria; for the LDA criterium, the EPR correspond to the (dark/light) blue colored 
regions, and for the GMM criterium, they correspond to the 95\% confidence intervals (ellipses). 
Fig. \ref{LDA_and_GMM} illustrates them clearly. Thus, when $\lambda_{A} > \lambda_{B}$ , the 
trajectory will spend, on average, more time in the EPR for A than 
in the EPR for B (and vice-versa when $\lambda_{B} > \lambda_{A}$). 

\begin{figure}[H]
\centering
\includegraphics[width=0.45\textwidth]{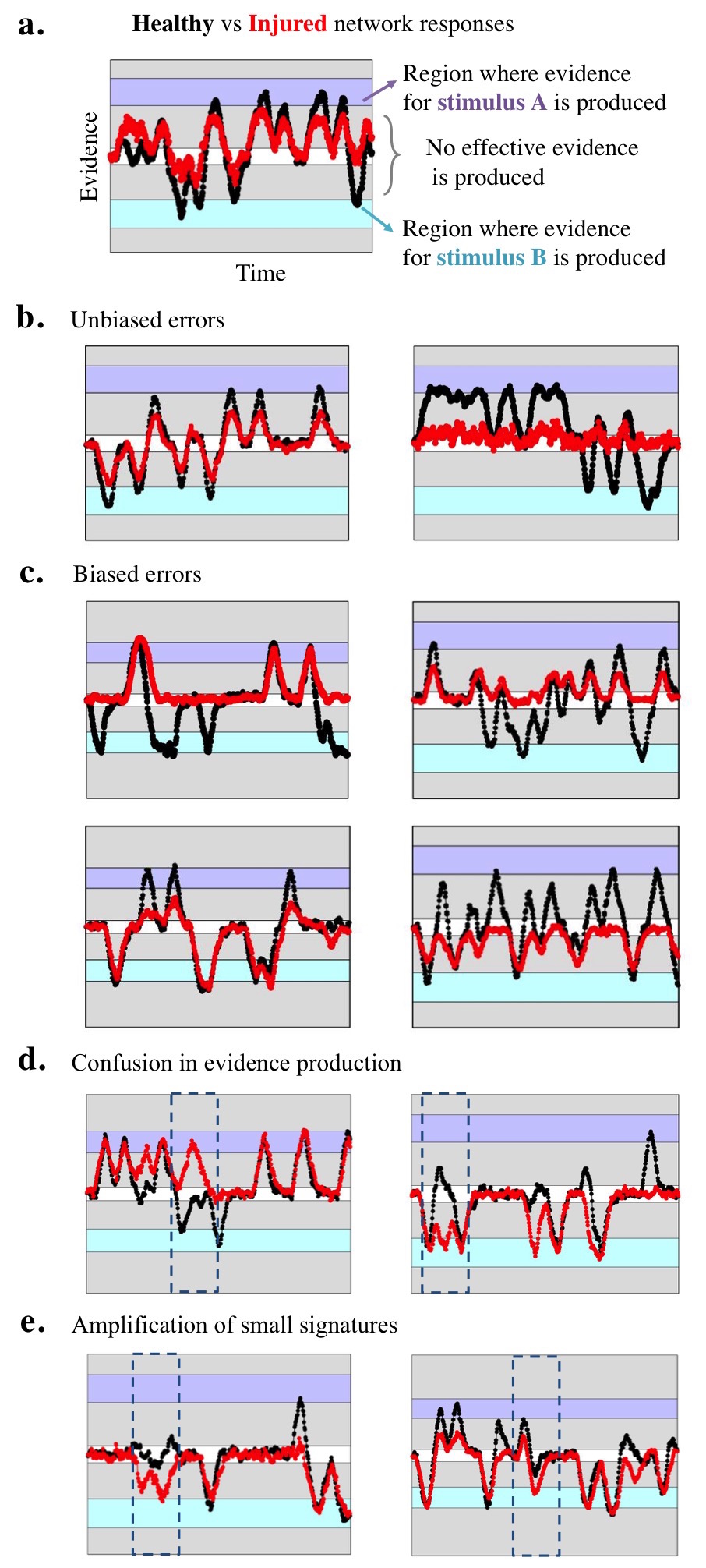}
\caption{Anomalous production of evidence. 
~ \textbf{a.} Comparison of evidence production in a healthy/injured 
neuronal network when the same stimuli is presented. 
 Colored regions on the top/bottom of the plot indicate which one of 
 the binary options for decision making (A vs B) the produced evidence 
 accounts for. 
 The injured response (red) may vary significantly from the healthy one (black). 
~ \textbf{b.} Examples of unbiased errors (when both options are 
penalized/jeopardized). 
~\textbf{c.} Examples of biased errors, i.e., when only one option is impaired 
or significantly more penalized than the other.
~\textbf{d.} Examples of confusion in evidence production (production of 
opposing/contrary evidence). 
~\textbf{e.} Amplification of small signatures and production of inexistent 
(false) evidence. 
The anomalous production of evidence will culminate in accuracy and speed 
deficits for the overall decision making process.} 
\label{anom_evid}       
\end{figure}

To map the trajectory from a 2-dimensional PCA plane to a (1D, diffusion-to-boundary) 
decision-making framework, we set 
\begin{equation}
\text{Evidence}(t) =  \big [ d_{B}(t) - d_{A}(t) \big ] / \big [ d_{B}(t) + d_{A}(t) \big ] ~,
\label{evidence}
\end{equation}
where $d_{A}$ and $d_{B}$ denote the euclidian distances between the trajectory and 
the fixed points $A$ and $B$ respectively. 
With this definition, the evidence values for choices $\{A,B\} $ will be approximately $\{+1,-1\}$
since $d_{A} \approx 0$ at the EPR for A and $d_{B} \approx 0$ at the EPR for B. When the 
network is at rest, and whenever $d_{A} \approx d_{B}$, the produced evidence will be 
small.  See Figure \ref{anom_evid}a  for a prototypical example and schematics. 
Notice that the EPR regions for choices A and B are mapped onto upper and lower 
rectangular bands respectively in the 1-dimensional plots. 

We discuss bellow a few  different type 
of anomalies regarding the production of evidence in injured networks. 
See Figure \ref{anom_evid}b--e for a tentative list of errors drawn from 
several injury levels and different FAS distributions.

\paragraph{Unbiased vs Biased errors:} Fig. \ref{anom_evid}b shows two examples of
injured network responses that produce unbiased errors, i.e., where the produced evidence 
jeopardizes both choices with similar magnitude. In the first plot, the injured response 
(in red) is slightly attenuated but still follows the healthy response (in black) closely. 
This occurs both when the signal is moving upwards (towards the EPR for choice A) and 
moving downwards (towards the EPR for choice B). 
In the second plot, the injured signal is heavily mitigated and produces no evidence for either choice.
This type of error contrasts with the ones depicted in Fig. \ref{anom_evid}c. For those plots, 
it is clear that one choice is significantly more penalized than the other. The upper plots 
in Fig. \ref{anom_evid}c still produce evidence for choice A (with different degrees of success), 
while the lower plots show analogous biases towards choice B. 
\paragraph{Confusion in evidence production:} Fig. \ref{anom_evid}d depicts 
two examples of a less common but significantly more dramatic type of error. 
Notice that in the first highlighted box of Fig. \ref{anom_evid}d, the healthy 
signal is clearly in the EPR for choice B while the injured one is in the EPR for choice A. 
Thus, an injured network may not only fail to produce evidence to a given choice but 
instead, produces evidence for the \textit{opposite} one.
A similar effect occurs to a lesser 
degree on the second plot. We believe that this unprecedented confusion mechanism 
operating at the evidence production level can lead to critical decision-making deficits.  

%
%

%
%
%
\begin{figure*}[th]
\centering
\includegraphics[width=0.9\textwidth]{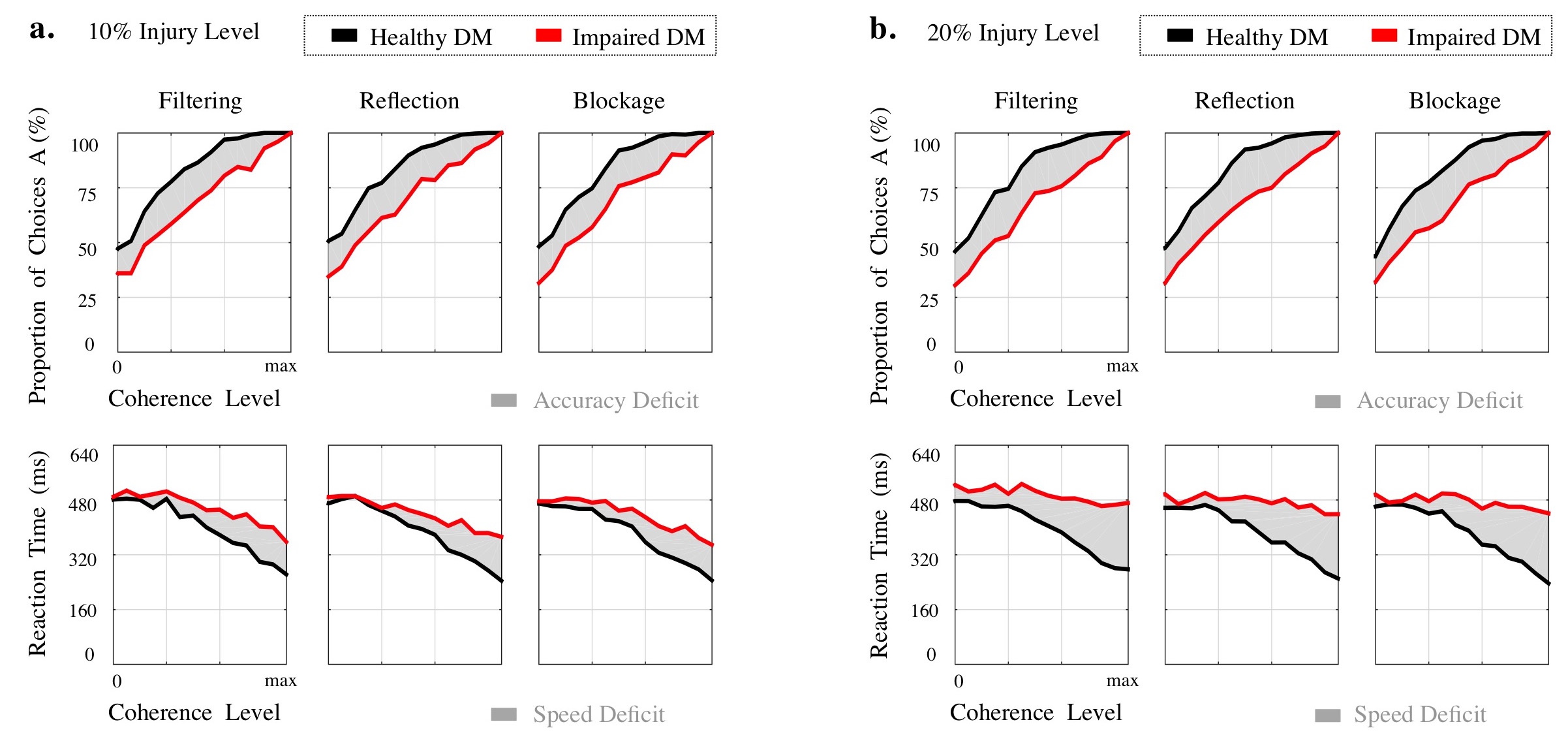}
\caption{Decision-Making impairments associated with injured neuronal networks. Binary choice 
task (A vs B). Results for homogenous FAS distributions, i.e., when all injured neurons display the 
same dysfunctional regime (filtering, reflection or blockage). ~\textbf{a.} Upper plots show the
Proportion of Choices A (\%) as a function of the coherence level (for A) within the presented 
stimulus;  zero coherence denotes a 50/50 Poisson mixed train of  A and B pulses while max. 
coherence only includes A pulses. Performance deficits are shaded in gray and highlight the 
difference between the healthy responses (in black) and networks injured at a 10\% level 
(in red). Lower plots compare the average time required to make the decision at 
each coherence level. ~\textbf{b.} 
Analogous plots for networks with 20\%-injured neuronal subpopulation. Doubling the fraction of injured
neurons worsens the reaction time of the system significantly but keeps the accuracy deficits 
at similar levels. In every plot, we average over 400 trials at 15 different coherence levels. 
We also randomize the neurons targeted with FAS and the external train of stimulus every turn.}
\label{FRB_curves}       
\end{figure*}
%
%
%
%
%

\paragraph{Amplification of small signatures:} In Fig. \ref{anom_evid}e, we show 
injured networks producing evidence when the original healthy network does not. 
By inspecting the two highlighted boxes, we conjecture that FAS injuries or noise 
mechanisms may amplify small fluctuations, creating non-existent pieces of evidence. This 
type of error occurs sometimes in conjunction with other ones, such as those in Panels 
b-c for example. 

Figure \ref{anom_evid} should not be regarded as a systematic catalogue of all possible types 
of errors regarding the anomalous production of evidence. Instead, our goal was to simply 
illustrate the rich variety of injured responses generated from random, biophysically motivated 
injuries applied to our simulations.
Some anomalies, like those presented in Fig. \ref{anom_evid}d--e for example, are non-obvious, 
while the ones in Fig. \ref{anom_evid}b--c are more frequent. Our next step is to investigate the 
average effects of the anomalous production of evidence to the decision-making capabilities of the 
system.  Indeed, the statistical appearance of these various evidence producing
regimes are what drive compromised DM.

%
%
\subsection{Impaired decision making and cognitive deficits}
%
%

%
%
%
\begin{figure*}[th]
\centering
\includegraphics[width=0.85\textwidth]{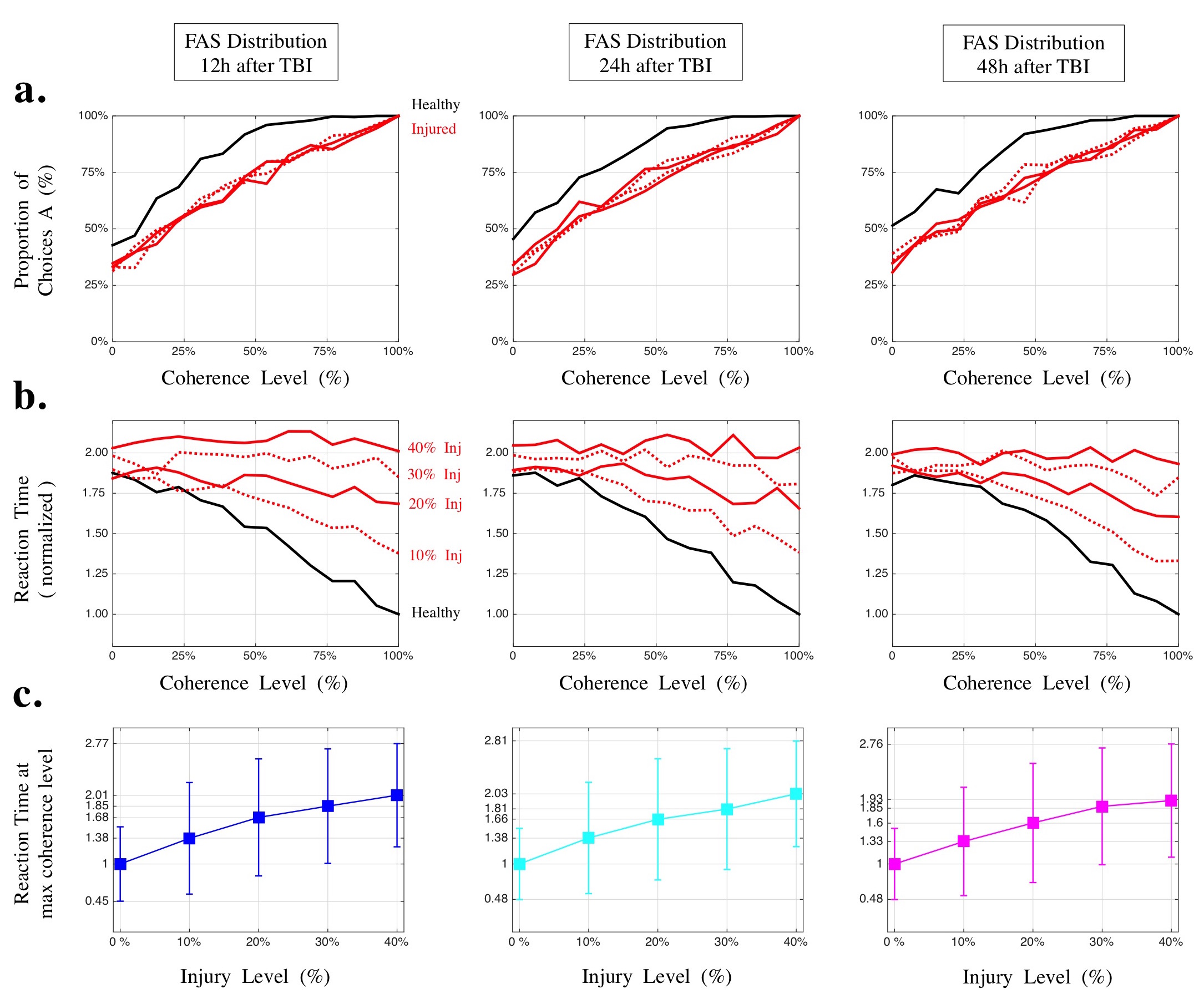}
\caption{Decision-Making impairments associated with injured neuronal networks. 
Binary choice task (A vs B). The FAS distributions were derived from Wang et al. \cite{Wang2011}
and correspond to swelling development along the optic nerve of an adult rat 12h, 24h, and 
48h after TBI. Healthy responses are plotted in black and compared with networks injured at 
different levels (10\%, 20\%, 30\% and 40\%, depicted in red).
~\textbf{a.} Upper plots show the Proportion of Choices A (\%) as a function of the 
coherence level (for A) within the presented stimulus;  zero coherence denotes a 
50/50 Poisson mixed train of  A and B pulses while max. coherence only includes 
A pulses.  
~\textbf{b.} The middle plots compare the average time required to make the decision at 
different coherence and injury levels. We normalized the system's reaction time using 
the healthy value at the maximum coherence level as a baseline. Within each plot, we 
see distinctive trends for different injury levels. 
~\textbf{c.} The bottom plots show the reaction times ($\pm$ 1 std.)  at the maximum coherence 
level, where the separability between injury levels (\%) is highest. The similarity between the 
trends across all three FAS distributions is staggering, suggesting that speed deficits increase 
with injury at a robust rate. 
}
\label{wang_curves}       
\end{figure*}
%
%
%

We now couple the network's mixed responses to stimuli with an accumulation 
model that ultimately decides in favor of Choice $\bf{A}$ or $\bf{B}$. Specifically, 
we assume (as in Shadlen et al. \cite{Shadlen2007}) that a \textit{Decision Variable} 
(DV) continuously integrates the produced evidence until a decision-boundary is reached. 
Then, the process is interrupted and the DV commits to that choice. When one 
of the alternatives dominates the mixed stimulus (high coherence level), the DV promptly 
reaches the correct decision boundary.   For low coherence, the decision typically
happens over a long period of time and is often incorrect.

The behavioral responses associated with different difficulty levels 
are well described by psychophysical curves,  i.e., by regarding the system's accuracy 
and reaction time as a function of the coherence level. See \cite{Shadlen2007} for an 
introductory exposition of the theme. In what follows, we present a series of simulations
where FAS hinder the neuronal network activity. As suggested by the previous
sections, the anomalous production of evidence leads to several deficits, but most importantly, 
to slower reaction times.

\paragraph{Homogeneous FAS distributions:} In Figure \ref{FRB_curves}, we show
psychophysical curves associated with injured networks in which all the FAS have the 
same functional type (filtering, reflection or blockage). 
The upper plots in Panel \ref{FRB_curves}a show how the proportion of choices $\bf{A}$ 
increases with the coherence level (for A). There is an accuracy deficit (shaded in gray) 
between the healthy system (in black) and the 10\% injured system (in red) for all three cases.
The type of FAS did not seem to strongly influence the results. 
The lower plots in Panel \ref{FRB_curves}a show psychometric curves for reaction times 
as a function of the coherence level. These curves are naturally decreasing since a 
higher coherence level is associated with an  easier task, and consequently, with faster 
decisions. Still, one can observe reaction time deficits across all homogeneous FAS distributions. 

Panel  \ref{FRB_curves}b  shows analogous plots for networks injured at a 20\% level. 
In every plot, we average over 400 trials at 15 different coherence levels, from zero 
($\lambda_{A} = \lambda_{B}$) to fully coherent ($\lambda_{A} \gg \lambda_{B} = 0$). 
We also randomize the neurons targeted with FAS and the external train of stimulus every 
turn. Overall, the impairments translate into less accurate and slower decision-making 
capabilities, although reaction time impairments are more pronounced at higher injury levels.

\paragraph{FAS distributions from Wang et al. \cite{Wang2011}:} Figure \ref{wang_curves} 
shows the central result of our set of simulations. The neuronal networks are now injured following the
FAS distributions derived from the experimental results of Wang et al. \cite{Wang2011}.
Each column corresponds to FAS development along the optic nerve of an adult rat  at 12h, 24h,  
and 48h after TBI. See Section \ref{infer_fas_param} and Fig. \ref{FAS_and_PIES}  for details.
In Panel \ref{wang_curves}a we show how the proportion of choices $\bf{A}$ increases with the 
coherence level (for A). Again, there is an accuracy deficit between the healthy responses (in black) 
and the injured ones (in red). The trends are similar for all three FAS distributions. 

\paragraph{Reaction-time impairments and network diagnostics:}
Panel \ref{wang_curves}b shows psychometric curves for the Reaction Times (RT).
Within each plot, we see distinctive trends for different injury levels (10\%, 20\%, 30\% and 40\% of 
damaged neurons in the network). 
We normalized the RT using the healthy value at the maximum coherence level 
as a baseline (i.e., setting RT = 1 when $\lambda_{A} \gg \lambda_{B} = 0$). The similarity 
between the trends across all three FAS distributions is staggering, suggesting that reaction time 
deficits increase with injury at a significant rate. 
Panel \ref{wang_curves}c and Table \ref{wang_table} show the RT ($\pm$ 1 std.) at 
the maximum coherence level, where the separability between injury levels (\%) is highest.  

The compromised reaction times have profound translational implications, specifically because
reaction times can be measured experimentally in a non-invasive fashion.
Thus, if a TBI patient takes 
(on average) around two times longer to successfully make a simple decision, our results 
suggest that roughly 40\% of the neurons in the network could have some form of FAS injury. 
More importantly, Table \ref{wang_table} could provide an indirect way to diagnose and calibrate FAS progression from clinically 
observed functional deficits.

%
%
%
\begin{table}
\centering
\caption{Reaction Time at max. coherence value for 
different injury levels. Healthy response time used as unit.
FAS distributions developed 12h, 24h, and 48h after TBI 
were inferred from Wang et al. \cite{Wang2011}.
  }
\begin{tabular}{c | c | c  | c }
\text{Inj. Level} & 12h & 24h & 48h \\
\hline
0   \% & 1.00 $\pm$ .54  & 1.00 $\pm$ .52 & 1.00 $\pm$ .52 \\
10 \% & 1.37 $\pm$ .81 & 1.38 $\pm$ .81 & 1.33 $\pm$ .79 \\
20 \% & 1.68 $\pm$ .86 & 1.65 $\pm$ .89 & 1.60 $\pm$ .87 \\
30 \% & 1.85 $\pm$ .84 & 1.80 $\pm$ .88 & 1.84 $\pm$ .85 \\
40 \% & 2.01 $\pm$ .75 & 2.03 $\pm$ .77 & 1.93 $\pm$ .83 \\
\end{tabular}
\label{wang_table}
\end{table}
%
%
%

%
%
%
%
%
%
\section{Discussion and Summary of Model}
\label{Discussion}
In this work, we develop a theoretical framework to quantify Decision-Making (DM) 
impairments following Traumatic Brain Injuries (TBI) and/or under the influence of neurodegenerative diseases. 
Specifically, we were able to link a cellular-level pathology named Focal Axonal Swellings (FAS) 
to network dysfunctions, to the anomalous production of evidence, and consequently, to DM 
impairments and cognitive deficits; see Fig. \ref{intro_scheme} for a simplified overview of 
our work. We introduced the key elements of our multi-scale model gradually throughout the 
paper, which we now summarize while revisiting Fig \ref{DM_scheme}:

\begin{enumerate}[(i)]
\item The decision-maker in Panel \ref{DM_scheme}A is faced with a \textit{binary choice} (A vs B)
and stimulated  by a random mixed train of $\bold{S_{A}}$ and $\bold{S_{B}}$ pulses 
(see eqs. \ref{train_stim} -- \ref{train_stim2}).
These pulses are generated by \textit{Poisson processes} of intensities $\lambda_{A}$ and 
$\lambda_{B}$, that are adjusted to produce different coherence levels (as in Panel \ref{DM_scheme}B). 
\\
\item The core of the DM system is a neuronal network that responds to 
meaningful stimuli with robust \textit{population codes}. In a proper low dimensional PCA space, 
input stimulus dynamically evolve the population codes \textit{fixed points} associated with each stimulus.  
Due to stochastic fluctuations,  trajectories fluctuate around the fixed points and produce evidence for
that stimulus. See Fig. \ref{Net_&_Traj} for details.
\\
\item The network responds to mixed stimuli with a series of trajectories that alternate 
between the vicinities of the fixed points A and B -- treated as \textit{Evidence-Producing-Regions} 
(see Fig. \ref{LDA_and_GMM}).  
Equation \ref{evidence} maps the trajectories to a 1D diffusion-to-boundary framework, 
and Fig. \ref{anom_evid} has an analogous role to Panel \ref{DM_scheme}C.
\\
\item The \textit{produced evidence} is coupled to an accumulation model; as in Panel 
\ref{DM_scheme}D, a \textit{decision variable} continuously integrates evidence, and commits 
to a choice when it reaches a boundary associated to that alternative. 
\\
\item The system displays plausible \textit{psychophysical curves} of accuracy/speed as a function 
of the coherence level (see Figs. \ref{FRB_curves} and \ref{wang_curves}). 
\end{enumerate}

In Section \ref{Background}, we provide a comprehensive review of FAS, a key signature 
of TBI reported in almost every animal study or \textit{in vitro} experiment. FAS are also 
present in Alzheimer's disease, which is the most common type of 
dementia in the elderly population, and several other neural disorders such as Multiple Sclerosis 
and Parkinson's disease. See Sec. \ref{Background} for a large list of references. 
The recent results of Maia and Kutz described several deleterious effects of FAS to spike propagation
\cite{Maia2015,Maia2014_2,Maia2014_1}. 
In fact, they demonstrated that the morphology (shape) of a swelling plays a critical role and can lead to 
the filtering, reflection, and blockage of action potentials. See Fig. \ref{spike_scheme}. 
In Sec. \ref{Methods}, we infer FAS
distributions from the experimental work of Wang et al. \cite{Wang2011} and include them in our 
simulations, which led to a variety of novel results:

\begin{enumerate}[(i)]
\item FAS hinders the network's production of low-dimensional population codes from their associated stimulus. Several trajectories 
no longer reach the expected fixed points and fluctuate around them. See Figs. \ref{Net_&_Traj} 
and \ref{LDA_and_GMM}. Confusion matrices show that the system's responses to meaningful 
stimuli become more random.
\\
\item There is a loss of Transmitted Information (TI) in the network. Figure \ref{Loss_Info} shows
the decay rate of TI  as a function of the injury level for different FAS distributions, including homogeneous 
FAS distributions and the ones inferred from Wang et al. \cite{Wang2011}.
\\
\item Distortion of population codes leads to anomalous production of evidence. There 
is significant variability in this aspect; Fig. \ref{anom_evid} shows examples of biased/unbiased errors, 
confusion (i.e., production of the wrong type of evidence), and amplification of (false) small signatures. 
\\
\item The anomalous production of evidence leads to DM impairments, which are easily interpretable 
from psychophysical curves for accuracy and speed. See Figs. \ref{FRB_curves} and \ref{wang_curves}.
We observe distinctive trends for \textit{reaction time impairments} as a function of the injury level, which could lead 
to novel FAS diagnostics \textit{in vivo}.
\end{enumerate}

Overall, we provided an innovative translational approach to bridge the large body of FAS literature
with the state of the art computational modeling of network dynamics and decision-making. The study of 
biologically-plausible FAS injuries and their pathological effects in computational neuroscience is
still in its infancy. Nevertheless, it is a direction for future research that could bring important insights to
our understanding of traumatic brain injuries and several neurodegenerative diseases.  
%
%
%
%
%
%
\section{Outlook}

A central finding of this work concerns reaction time impairments in DM networks.  This 
result has significant translational value to practitioners since it can be measured in practice.
As such, the outlook of this work has two distinct applications, one for neurodegenerative diseases
and a second for TBI.  They are characterized in what follows.

\subsection{Neurodegeneration and disease progression}

Unlike TBI, which has substantial experimental evidence supporting the potential reduction in FAS over time in
a neuronal network, neurodegenerative diseases are progressive over time and simply continue to 
destroy neurons in the network.  In addition to attempts to arrest the spread of the disease, assessment
of its progression over time is critical.  The findings here suggest that patients with neurodegenerative 
diseases can be given simple cognitive and/or motor tests, for instance, on tablet computers, to assess
their reaction times.  By monitoring their increasing reaction times to such tests, a readily interpretable
framework can be used to assess the damage level, or spread of FAS, in the neuronal network.  The
assessment is non-invasive, and can help determine the rate of progression of the disease so that 
personalized treatments can be made for patients with rapid or slow spread of disease.

In addition, the tests are predictive and can be used to potentially predict the loss of key
cognitive and motor functions, especially when considering the course of disease spread through
various brain regions and along the spinal cord.  
Of course, continued behavioral and cognitive testing would be required to produce this level of 
diagnostic, but the theoretical framework developed here provides a mathematical architecture
that can potentially help inform such predictive pathways for deterioration of cognitive and
motor function.

\subsection{TBI and concussion metrics}

The Glasgow Coma Scale (GCS) is a common test used by First Aid responders to assess the 
level of consciousness of a person after head injury. It is composed of three parts 
(eye, verbal and motor), where the individual values (from 1 to 5) are recorded and summed. 
The lowest summed score (GCS = 3) corresponds to deep coma or 
death, while the highest score (GCS = 15) corresponds to a fully awake person. 
The brain injury is then classified as severe (9 or less), moderate (9-12) or 
minor (13 and higher). 

Unfortunately, many contact-sport practitioners experience 
mild TBI (or concussions) that are rarely or only vaguely examined by the time 
they occur. To make matters worse, our results suggest that even a highly injured 
neuronal network could still produce somewhat accurate responses to simple tasks. 
This feature could easily mislead the hasty medical examinations given to athletes during 
(or after) a match. Our main results (Fig. \ref{wang_curves} and Table \ref{wang_table})
indicate that we should instead, focus on \textit{reaction time deficits} even when correct
answers are given. Thus, we recommend that athletes complete an array of quick, 
tablet-based games before matches to use as a baseline for normal reaction times. 
Once a potential head injury occurs, they would perform those same tasks and have a
software compare their performance. The software should have an ample selection of 
games to overcome the large variability of impairments that could occur on a single 
decision-making task.

%
%
%
%
%
%


%
%


\begin{thebibliography}{10}
\providecommand{\url}[1]{{#1}}
\providecommand{\urlprefix}{URL }
\expandafter\ifx\csname urlstyle\endcsname\relax
  \providecommand{\doi}[1]{DOI~\discretionary{}{}{}#1}\else
  \providecommand{\doi}{DOI~\discretionary{}{}{}\begingroup
  \urlstyle{rm}\Url}\fi

\bibitem{Adalbert2009}
Adalbert, R., Nogradi, A., Babetto, E., Janeckova, L., Walker, S.A.,
  Kerschensteiner, M., Misgeld, T., Coleman, M.P.: Severely dystrophic axons at
  amyloid plaques remain continuous and connected to viable cell bodies.
\newblock BRAIN \textbf{132}, 402--416 (2009)

\bibitem{Adams2011}
Adams, J.H., Jennett, B., Murray, L.S., Teasdale, G.M., Gennarelli, T.A.,
  Graham, D.I.: Neuropathological findings in disabled survivors of a head
  injury.
\newblock Journal of Neurotrauma \textbf{28}, 701--709 (2011)

\bibitem{Adle1999}
Adle-Biassette, H., Chretien, F., Wingertsmann, L., Hery, C., Ereau, T.,
  Scaravilli, F., Tardieu, M., Gray, F.: Neuronal apoptosis does not correlate
  with dementia in hiv infection but is related to microglial activation and
  axonal damage.
\newblock Neuropathology and Applied Neurobiology \textbf{25}, 123--133 (1999)

\bibitem{Barnes2014}
Barnes, D.E., Kaup, A., Kirby, K., Byers, A.L., R.Diaz-Arrastia, Yaffe, K.:
  Traumatic brain injury and risk of dementia in older veterans.
\newblock Neurology \textbf{83}, 312--319 (2014)

\bibitem{Blumbergs1995}
Blumbergs, P., Scott, G., Manavis, J., Wainwright, H., Simpson, D., McLean, A.:
  Topography of axonal injury as defined by amyloid precursor protein and the
  sector scoring method in mild and severe closed head injury.
\newblock Journal of Neurotrauma \textbf{12}, 565--572 (1995)

\bibitem{Bogacz2006}
Bogacz, R., Brown, E., Moehlis, J., Holmes, P., Cohen, J.D.: The physics of
  optimal decision making: A formal analysis of models of performance in
  two-alternative forced-choice tasks.
\newblock Psychological Review \textbf{113}(4), 700--765 (2006)

\bibitem{Network_Review}
Brette, R., Rudolph, M., Carnevale, T., Hines, M., Beeman, D., Bower, J.M.,
  Diesmann, M., Morrison, A., Goodman, P.H., Jr., F.C.H., Zirpe, M.,
  Natschläger, T., Pecevski, D., Ermentrout, B., Djurfeldt, M., Lansner, A.,
  Rochel, O., Vieville, T., Muller, E., Davison, A.P., Boustani, S.E.,
  Destexhe, A.: Simulation of networks of spiking neurons: A review of tools
  and strategies.
\newblock Journal of Computational Neuroscience \textbf{23}(3), 349--398 (2007)

\bibitem{Browne2011}
Browne, K.D., Chen, X.H., Meaney, D.F., Smith, D.H.: Mild traumatic brain
  injury and diffuse axonal injury in swine.
\newblock Journal of Neurotrauma \textbf{28}(9), 1747--1755 (2011)

\bibitem{Brunton2013}
Brunton, B.W., Botvinick, M.M., Brody, C.D.: Rats and humans can optimally
  accumulate evidence for decision-making.
\newblock Science \textbf{340}, 95--98 (2013)

\bibitem{Chen2009}
Chen, Y.C., Smith, D.H., Meaney, D.: In-vitro approaches for studying
  blast-induced traumatic brain injury.
\newblock Journal of Neurotrauma \textbf{26}(6), 861--876 (2009)

\bibitem{Christman1994}
Christman, C., Grady, M., Walker, S., Hol-Loway, K., Povlishock, J.:
  Ultra-structural studies of diffuse axonal injury in humans.
\newblock Journal of Neurotrauma \textbf{11}, 173--186 (1994)

\bibitem{Coleman2005}
Coleman, M.: Axon degeneration mechanisms: commonality amid diversity.
\newblock Nature Reviews Neuroscience \textbf{6}(11), 889--898 (2005)

\bibitem{Daianu2016}
Daianu, M., Jacobs, R.E., Town, T., Thompson, P.M.: Axonal diameter and density
  estimated with 7-tesla hybrid diffusion imaging in transgenic alzheimer rats.
\newblock SPIE Proceedings \textbf{9784}, 1--6 (2016)

\bibitem{Dayan2001}
Dayan, P., Abbot, L.: Theoretical neuroscience.
\newblock MIT Press (2001)

\bibitem{Dikranian2008}
Dikranian, K., Cohen, R., Donald, C.M., Pan, Y., Brakefield, D., Bayly, P.,
  Parsadanian, A.: Mild traumatic brain injury to the infant mouse causes
  robust white matter axonal degeneration which precedes apoptotic death of
  cortical and thalamic neurons.
\newblock Experimental Neurology \textbf{211}, 551--560 (2008)

\bibitem{Ditterich2006}
Ditterich, J.: Stochastic models of decisions about motion direction: Behavior
  and physiology.
\newblock Neural Networks \textbf{19}, 981--1012 (2006)

\bibitem{Edlow2016}
Edlow, B.L., Copen, W.A., Izzy, S., van~der Kouwe, A., Glenn, M.B., Greenberg,
  S.M., Greer, D.M., Wu, O.: Longitudinal diffusion tensor imaging detects
  recovery of fractional anisotropy within traumatic axonal injury lesions.
\newblock Neurocritical Care \textbf{24}(3), 342--352 (2016)

\bibitem{book_SI}
Fainaru-Wada, M., Fainaru, S.: League of denial: The nfl, concussions, and the
  battle for truth.
\newblock Crown Archetype  (2013)

\bibitem{cdc}
Faul, M., Xu, L., Wald, M.M., Coronado, V.G.: Traumatic brain injury in the
  united states: emergency department visits, hospitalizations, and deaths.
\newblock Atlanta (GA): Centers for Disease Control and Prevention, National
  Center for Injury Prevention and Control  (2010)

\bibitem{Fayaz2000}
Fayanz, I., Tator, C.H.: Modeling axonal injury in vitro: injury and
  regeneration following acute neuritic trauma.
\newblock Journal of Neuroscience Methods \textbf{102}, 69--79 (2000)

\bibitem{Friese2014}
Friese, M.A., Schattling, B., Fugger, L.: Mechanisms of neurodegeneration and
  axonal dysfunction in multiple sclerosis.
\newblock Nature Reviews Neurology \textbf{10}, 225--238 (2014)

\bibitem{Galvin1999}
Galvin, J.E., Uryu, K., Lee, V.M., Trojanowski, J.Q.: Axon pathology in
  parkinsonÕs disease and lewy body dementia hippocampus contains $\alpha$-,
  $\beta$-, and $\gamma$ -synuclein.
\newblock Proceedings of National Academy of Science \textbf{96},
  13,450--13,455 (1999)

\bibitem{Grady1993}
Grady, M., Mclaughlin, M., Christman, C., Valadaka, A., Flinger, C.,
  Povlishock, J.: The use of antibodies against neurofilament subunits for the
  detection of diffuse axonal injury in humans.
\newblock Journal of Neuropathology and Experimental Neurology \textbf{52},
  143--152 (1993)

\bibitem{Gupta2016}
Gupta, R., Sen, N.: Traumatic brain injury: a risk factor for neurodegenerative
  diseases.
\newblock Reviews in the Neurosciences \textbf{27}(1), 93?100 (2016)

\bibitem{Hanell2015}
Hanell, A., Greer, J.E., McGinn, M.J., Povlishock, J.T.: Traumatic brain
  injury?induced axonal phenotypes react differently to treatment.
\newblock Acta Neuropathologica \textbf{129}, 317--332 (2015)

\bibitem{Hay2016}
Hay, J., Johnson, V.E., Smith, D.H., Stewart, W.: Chronic traumatic
  encephalopathy: the neuropathological legacy of traumatic brain injury.
\newblock Annual Review of Pathology: Mechanisms of Disease \textbf{11}, 21--45
  (2016)

\bibitem{Hellman2010}
Hellman, A.N., Vahidi, B., Kim, H.J., Mismar, W., Steward, O., Jeonde, N.L.,
  Venugopalan, V.: Examination of axonal injury and regeneration in
  micropatterned neuronal culture using pulsed laser microbeam dissection.
\newblock Lab on a Chip \textbf{16}, 2083Ð2092 (2010)

\bibitem{Hemphill2011}
Hemphill, M., Dabiri, B., Gabriele, S., Kerscher, L., Franck, C., Goss, J.,
  Alford, P., Parker, K.: A possible role for integrin signaling in diffuse
  axonal injury.
\newblock PLos ONE \textbf{6}(7), e22,899 (2011)

\bibitem{Hemphill2015}
Hemphill, M., Dauth, S., Yu, C.J., Dabiri, B., Parker, K.: Traumatic brain
  injury and the neuronal microenvironment: A potential role for
  neuropathological mechanotransduction.
\newblock Neuron \textbf{86}(6), 1177--1192 (2015)

\bibitem{Henninger2016}
Henninger, N., Bouley, J., Sikoglu, E.M., An, J., Moore, C.M., King, J.A.,
  Bowser, R., Freeman, M.R., Jr, R.H.B.: Attenuated traumatic axonal injury and
  improved functional outcome after traumatic brain injury in mice lacking
  sarm1.
\newblock BRAIN pp. 1--12 (2016)

\bibitem{Herwerth2016}
Herwerth, M., Kalluri, S.R., Srivastava, R., Kleele, T., Kenet, S., Illes, Z.,
  Merkler, D., Bennett, J.L., Misgeld, T., Hemmer, B.: In vivo imaging reveals
  rapid astrocyte depletion and axon damage in a model of neuromyelitis
  optica-related pathology.
\newblock Annals of Neurology \textbf{79}, 794--805 (2016)

\bibitem{Higham2001}
Higham, D.: An algorithmic introduction to numerical simulation of stochastic
  differential equations.
\newblock SIAM Review \textbf{43}(3), 525 (546)

\bibitem{Hill2016}
Hill, C.S., Coleman, M.P., Menon, D.K.: Traumatic axonal injury: mechanisms and
  translational opportunities.
\newblock Trends in Neuroscience \textbf{39}(5), 311--324 (2016)

\bibitem{Ikonomovic2004}
Ikonomovic, M.D., Uryu, K., Abrahamson, E.E., Ciallella, J.R., Trojanowski,
  J.Q., Lee, V.M.Y., Clark, R.S., Marione, D.W., Wisniewski, S.R., DeKosky,
  S.T.: Alzheimer?s pathology in human temporal cortex surgically excised after
  severe brain injury.
\newblock Experimental Neurology \textbf{190}, 192--203 (2004)

\bibitem{Johnson2010}
Johnson, V.E., Stewart, W., Smith, D.H.: Traumatic brain injury and
  amyloid-$\beta$ pathology: a link to alzheimer's disease?
\newblock Nature Reviews Neuroscience \textbf{11}, 361--370 (2010)

\bibitem{Johnson2012}
Johnson, V.E., Stewart, W., Smith, D.H.: Widespread tau and amyloid-beta
  pathology many years after a single traumatic brain injury in humans.
\newblock Brain Pathology \textbf{22}, 142--149 (2012)

\bibitem{Johnson2013}
Johnson, V.E., Stewart, W., Smith, D.H.: Axonal pathology in traumatic brain
  injury.
\newblock Experimental Neurology \textbf{246}, 35--43 (2013)

\bibitem{jones07}
Jones, L., Fontanini, A., Sadacca, B., Miller, P., Katz, D.: Natural stimuli
  evoke dynamic sequences of states in sensory cortical ensembles.
\newblock Proc. Natl. Acad. Sci. USA \textbf{104}, 18,772--18,777 (2007)

\bibitem{Jorge2012}
Jorge, R.E., Acion, L., White, T., Tordesillas-Gutierrez, D., Pierson, R.,
  Crespo-Facorro, B., Magnotta, V.: White matter abnormalities in veterans with
  mild traumatic brain injury.
\newblock American Journal of Psychiatry \textbf{169}(12), 1284--1291 (2012)

\bibitem{Jorm1998}
Jorm, A.F., Jolley, D.: The incidence of dementia: a meta analysis.
\newblock Neurology \textbf{51}, 728--733 (1998)

\bibitem{Karlsson2016}
Karlsson, P., Haroutounian, S., Polydefkis, M., Nyengaard, J.R., Jensen, T.S.:
  Structural and functional characterization of nerve fibres in polyneuropathy
  and healthy subjects.
\newblock Scandinavian Journal of Pain \textbf{10}, 28--35 (2016)

\bibitem{Kinnunen2010}
Kinnunen, K.M., Greenwood, R., Powell, J.H., Leech, R., Hawkins, P.C.,
  Bonnelle, V., Patel, M.C., Counsell, S.J., Sharp, D.J.: White matter damage
  and cognitive impairment after traumatic brain injury.
\newblock Brain pp. 1--15 (2010)

\bibitem{Kolaric2013}
Kolaric, K.V., Thomson, G., Edgar, J.M., Brown, A.M.: Focal axonal swellings
  and associated ultrastructural changes attenuate conduction velocity in
  central nervous system axons: a computer modeling study.
\newblock Physiological reports \textbf{1}(3), e00,059 (2013)

\bibitem{Krstic2012}
Krstic, D., Knuesel, I.: Deciphering the mechanism underlying late-onset
  alzheimer disease.
\newblock Nature Reviews Neuroscience \textbf{9}(1), 25--34 (2012)

\bibitem{Lachance2014}
Lachance, M., Longtin, A., Morris, C.E., Yu, N., Joós, B.: Stimulation-induced
  ectopicity and propagation windows in model damaged axons.
\newblock Journal of Computational Neuroscience \textbf{37}, 523--531 (2014)

\bibitem{Laukka2016}
Laukka, J.J., Kamholz, J., Bessert, D.: Novel pathologic findings in patients
  with pelizaeus-merzbacher disease.
\newblock Neuroscience Letters  (2016)

\bibitem{Lauria2003}
Lauria, G., Morbin, M., Lombardi, R., Borgna, M., Mazzoleni, G., Sghirlanzoni,
  A., Pareyson, D.: Axonal swellings predict the degeneration of epidermal
  nerve fibers in painful neuropathies.
\newblock Neurology \textbf{61}, 631--636 (2003)

\bibitem{Liberski1999}
Liberski, P.P., Budka, H.: Neuroaxonal pathology in creutzfeldt-jakob disease.
\newblock Acta Neuropathology \textbf{97}, 329--334 (1999)

\bibitem{Lobue2016}
LoBue, C., Denney, D., Hynan, L.S., Rossetti, H.C., Lacritz, L.H., Jr., J.H.,
  Womack, K.B., Woon, F.L., Cullum, C.M.: Self-reported traumatic brain injury
  and mild cognitive impairment: increased risk and earlier age of diagnosis.
\newblock Journal of Alzheimer?s Disease \textbf{51}, 727--736 (2016)

\bibitem{Louis2009}
Louis, E.D., Faust, P.L., Vonsattel, J., Honig, L.S., Rajput, A., Rajput, A.,
  Pahwa, R., Lyons, K.E., Ross, G.W., Elble, R.J., Erickson-Davis, C.,
  Moskowitz, C.B., Lawton, A.: Torpedoes in parkinson?s disease, alzheimer?s
  disease, essential tremor, and control brains.
\newblock Movement Disorders \textbf{24}(11), 1600--1605 (2009)

\bibitem{Magdesian2016}
Magdesian, M.H., Lopez-Ayon, G.M., Mori, M., Boudreau, D., Goulet-Hanssens, A.,
  Sanz, R., Miyahara, Y., Barrett, C.J., Fournier, A.E., Koninck, Y.D.,
  Gr{\"u}tter, P.: Rapid mechanically controlled rewiring of neuronal circuits.
\newblock The Journal of Neuroscience \textbf{36}(3), 979--987 (2016)

\bibitem{Magdesian2012}
Magdesian, M.H., Sanchez, F., Lopez, M., Thostrup, P., Durisic, N., Belkaid,
  W., Liazoghli, D., Gr\"{u}tter, P., Colman, R.: Atomic force microscopy
  reveals important differences in axonal resistance to injury.
\newblock Biophysical Journal \textbf{103}(3), 405--414 (2012)

\bibitem{Maia2015}
Maia, P.D., Hemphill, M.A., Zehnder, B., Zhang, C., Parker, K.K., Kutz, J.N.:
  Diagnostic tools for evaluating the impact of focal axonal swellings arising
  in neurodegenerative diseases and/or traumatic brain injury.
\newblock Journal of Neuroscience Methods \textbf{253}, 233--243 (2015)

\bibitem{Maia2014_2}
Maia, P.D., Kutz, J.N.: Compromised axonal functionality after
  neurodegeneration, concussion and/or traumatic brain injury.
\newblock Journal of Computational Neuroscience \textbf{27}, 317--332 (2014)

\bibitem{Maia2014_1}
Maia, P.D., Kutz, J.N.: Identifying critical regions for spike propagation in
  axon segments.
\newblock Journal of Computational Neuroscience \textbf{36}(2), 141--155 (2014)

\bibitem{Maxwell1997}
Maxwell, W.L., Povlishock, J.T., Graham, D.L.: A mechanistic analysis of
  nondisruptive axonal injury: A review.
\newblock Journal of Neurotrauma \textbf{17}(7), 419--440 (1997)

\bibitem{Menon2015}
Menon, D.K., Maas, A.I.R.: Progress, failures and new approaches for tbi
  research.
\newblock Nature Reviews Neuroloy \textbf{11}, 71--72 (2015)

\bibitem{Millecamps2013}
Millecamps, S., Julien, J.: Axonal transport deficits and neurodegenerative
  diseases.
\newblock Nature Reviews Neuroscience \textbf{14}(161), 161--176 (2013)

\bibitem{Morrison2011}
Morrison, B., Elkin, B.S., Dolle, J.P., Yarmush, M.L.: In vitro models of
  traumatic brain injury.
\newblock Annual Reviews in Biomedical Engineering \textbf{13}(1), 91--126
  (2011)

\bibitem{Nikic2011}
Nikic, I., Merkler, D., Sorbara, C., Brinkoetter, M., Kreutzfeld, M., Bareyre,
  F., Bruck, W., Bishop, D., Misgeld, T., Kerschensteiner, M.: A reversible
  form of axon damage in experimental autoimmune encephalomyelitis and multiple
  sclerosis.
\newblock Nature Medicine \textbf{17}(4), 495--499 (2011)

\bibitem{Park2013science}
Park, H.J., Friston, K.: Structural and functional brain networks: From
  connections to cognition.
\newblock Science \textbf{342}, 1238,411--1--1238,411--8 (2013)

\bibitem{Patterson2015}
Patterson, B.W., Elbert, D.L., Mawuenyega, K.G., Kasten, T., Ovod, V., Ma, S.,
  Xiong, C., Chott, R., Yarasheski, K., Sigurdson, W., Zhang, L., Goate, A.,
  Benzinger, T., Morris, J.C., Holtzman, D., Bateman, R.J.: Age and amyloid
  effects on human central nervous system amyloid-beta kinetics.
\newblock American Neurological Association \textbf{78}(3), 439--453 (2015)

\bibitem{Petersen2004}
Petersen, R.C.: Mild cognitive impairment as a diagnostic entity.
\newblock Journal of Internal Medicine \textbf{256}, 183--194 (2004)

\bibitem{Povlishock2005}
Povlishock, J.T., Katz, D.I.: Update of neuropathology and neurological
  recovery after traumatic brain injury.
\newblock Journal of Head Trauma Rehabilitation \textbf{20}(1), 76--94 (2005)

\bibitem{Qiu2009}
Qiu, C., Kivipelto, M., von Strauss, E.: Epidemiology of alzheimer?s disease:
  occurrence, determinants, and strategies toward intervention.
\newblock Dialogues in Clinical Neuroscience \textbf{11}(2), 111--128 (2009)

\bibitem{rabinovich08}
Rabinovich, M., Huerta, R., Varona, P., Afraimovich, V.: Transient cognitive
  dynamics, metastability, and decision making.
\newblock PLoS Comp. Bio. \textbf{4}, e1000,072 (2008)

\bibitem{rabinovich01}
Rabinovich, M., Volkovskii, A., Lecanda, P., Huerta, R., Abarbanel, H.,
  Laurent, G.: Dynamical encoding by networks of competing neuron groups:
  Winnerless competition.
\newblock Phys. Rev. Lettl. \textbf{87}, 068,102 (2001)

\bibitem{Ratcliff1998}
Ratcliff, R., Rouder, J.N.: Modeling response times for two-choice decisions.
\newblock Psychological Science \textbf{9}(5), 347--356 (1998)

\bibitem{Mau2016}
del Razo, M.J., Morofuji, Y., Meabon, J.S., Huber, B.R., Peskind, E.R., Banks,
  W.A., Mourad, P.D., LeVeque, R.J., Cook, D.G.: Computational and in vitro
  studies of blast-induced blood-brain barrier disruption.
\newblock SIAM Journal on Scientific Computing \textbf{38}(3), 347--374 (2016)

\bibitem{Reeves2012}
Reeves, T.M., Smith, T.L., Williamson, J.C., Phillips, L.L.: Unmyelinated axons
  show selective rostrocaudal pathology in the corpus callosum after traumatic
  brain injury.
\newblock Journal of Neuropathology \& Experimental Neurology \textbf{71}(3),
  198--210 (2012)

\bibitem{Riffell2014}
Riffell, J.A., Shlizerman, E., Sanders, E., Abrell, L., Medina, B.,
  Hinterwirth, A.J., Kutz, J.N.: Flower discrimination by pollinators in a
  dynamic chemical environment.
\newblock Science \textbf{344}, 1515--1518 (2014)

\bibitem{Roozenbeek2013}
Roozenbeek, B., Maas, A.I.R., Menon, D.K.: Changing patterns in the
  epidemiology of traumatic brain injury.
\newblock Nature Reviews Neurology \textbf{9}, 231--236 (2013)

\bibitem{Rubovich2011}
Rubovitch, V., Ten-Bosch, M., Zohar, O., Harrison, C., Tempel-Brami, C., Stein,
  E., Hoffer, B., Balaban, C., Schreiber, S., Chiu, W., Pick, C.: A mouse model
  of blast-induced mild traumatic brain injury.
\newblock Experimental Neurology \textbf{232}(2), 280--289 (2011)

\bibitem{Rudy2016}
Rudy, S., Maia, P.D., Kutz, J.N.: Cognitive and behavioral deficits arising
  from neurodegeneration and traumatic brain injury: a model for the underlying
  role of focal axonal swellings in neuronal networks with plasticity.
\newblock Journal of Systems and Integrative Neuroscience  (2016)

\bibitem{Shadlen2007}
Shadlen, M.N., Hanks, T.D., Churchland, A.K., Kiani, R., Yang, T.: The Speed
  and Accuracy of a Simple Perceptual Decision: A Mathematical Primer.
\newblock Ch.10 (2007)

\bibitem{Shadlen2013}
Shadlen, M.N., Kiani, R.: Decision making as a window on cognition.
\newblock Neuron \textbf{80}(3), 791--332 (2013)

\bibitem{Shadlen2016}
Shadlen, M.N., Shohamy, D.: Decision making and sequential sampling from
  memory.
\newblock Neuron \textbf{90}, 927--939 (2016)

\bibitem{Sharp2014}
Sharp, D.J., Scott, G., Leech, R.: Network dysfunction after traumatic brain
  injury.
\newblock Nature Reviews Neurology \textbf{10}, 156--166 (2014)

\bibitem{Shlizerman2014}
Shlizerman, E., Riffell, J.A., Kutz, J.N.: Data-driven inference of network
  connectivity for modeling the dynamics of neural codes in the insect antennal
  lobe.
\newblock Frontiers in Computational Neuroscience \textbf{8}(70), 1--15 (2014)

\bibitem{Skandsen2010}
Skandsen, T., Kvistad, K.A., Solheim, O., Strand, I.H., Folvik, M., Vik, A.:
  Prevalence and impact of diffuse axonal injury in patients with moderate and
  severe head injury: a cohort study of early magnetic resonance imaging
  findings and 1-year outcome.
\newblock Journal of Neurosurgery \textbf{113}(3), 556--563 (2010)

\bibitem{Smith1999}
Smith, D., Wolf, J., Lusardi, T., Lee, V., Meaney, D.: High tolerance and
  delayed elastic response of cultured axons to dynamic stretch injury.
\newblock The Journal of Neuroscience \textbf{19}(11), 4263--4269 (1999)

\bibitem{Tagliaferro2016}
Tagliaferro, P., Burke, R.E.: Retrograde axonal degeneration in parkinson
  disease.
\newblock Journal of Parkinson?s Disease \textbf{6}, 1--15 (2016)

\bibitem{TangSchomer2012}
Tang-Schomer, M.D., Johnson, V.E., Baas, P.W., Stewart, W., Smith, D.H.:
  Partial interruption of axonal transport due to microtubule breakage accounts
  for the formation of periodic varicosities after traumatic axonal injury.
\newblock Experimental Neurology \textbf{233}, 364--372 (2012)

\bibitem{TangSchomer2010}
Tang-Schomer, M.D., Patel, A., Bass, P.W., Smith, D.H.: Mechanical breaking of
  microtubules in axons during dynamic stretch injury underlies delayed
  elasticity, microtubule disassembly, and axon degeneration.
\newblock The FASEB Journal \textbf{24}(5), 1401--1410 (2010)

\bibitem{Thies2013}
Thies, W., Bleiler, L.: Alzheimer's disease facts and figures.
\newblock Alzheimer's \& Dementia \textbf{9}(2), 208?245 (2013)

\bibitem{Trapp2008}
Trapp, B.D., Nave, K.A.: Multiple sclerosis: An immune or neurodegenerative
  disorder?
\newblock Annual Review Neuroscience \textbf{31}(1), 247--269 (2008)

\bibitem{Tsai2004}
Tsai, J., Grutzendler, J., Duff, K., Gan, W.B.: Fibrillar amyloid deposition
  leads to local synaptic abnormalities and breakage of neuronal branches.
\newblock Nature Neuroscience \textbf{7}, 1181--1183 (2004)

\bibitem{Victor1997}
Victor, J.D., Purpura., K.P.: Metric space analysis of spike trains: theory,
  algorithms and application.
\newblock Network: Computational Neural Systems \textbf{8}, 127--164 (1997)

\bibitem{Wang2011}
Wang, J., Hamm, R.J., Povlishock, J.T.: Traumatic axonal injury in the optic
  nerve: evidence for axonal swelling, disconnection, dieback and
  reorganization.
\newblock Journal of Neurotrauma, \textbf{28}(7), 1185--1198 (2011)

\bibitem{Watts:1998db}
Watts, D.J., Strogatz, S.H.: {Collective dynamics of 'small-world' networks.}
\newblock Nature \textbf{393}(6684), 440--442 (1998)

\bibitem{Gerstner2014}
Wulfram~Gerstner Werner M.~Kistler, R.N., Paninski, L.: Neuronal Dynamics.
\newblock Cambridge University Press (2014)

\bibitem{Xiong2013}
Xiong, Y., Mahmood, A., Chopp, M.: Animal models of traumatic brain injury.
\newblock Nature Reviews Neuroscience \textbf{14}(22), 128--142 (2013)

\bibitem{Yue2013}
Yue, J.K., Vassar, M.J., Lingsma, H.F., Cooper, S.R., Okonkwo, D.O., Valadka,
  A.B., Gordon, W.A., Maas, A.I.R., Mukherjee, P., Yuh, E.L., Puccio, A.M.,
  Schnyer, D.M., Manley, G.T., Casey, S.S., Cheong, M., Dams-O?Connor, K.,
  Hricik, A.J., Knight, E.E., Kulubya, E.S., Menon, D.K., Morabito, D.J.,
  Pacheco, J.L., Sinha, T.K.: Transforming research and clinical knowledge in
  traumatic brain injury pilot: multicenter implementation of the common data
  elements for traumatic brain injury.
\newblock Journal of Neurotrauma \textbf{30}, 1831--1844 (2013)

\end{thebibliography}
\end{document}